%% file: sliding_grid.tex
\documentclass[10pt,a4paper]{article}

\usepackage[T1]{fontenc}
\usepackage[utf8]{inputenc}
\usepackage{amssymb,amsmath,amsthm,amsfonts}    
\usepackage{bm}                                 
\usepackage{mathtools}                          
\usepackage{stmaryrd}                           
\usepackage{breqn}                              
\usepackage[textfont=footnotesize,labelfont={footnotesize,bf}]{caption}
\usepackage{subcaption}
\usepackage{booktabs}
\usepackage[inline]{enumitem}
\usepackage{graphicx}
\usepackage[dvipsnames]{xcolor}
\usepackage[top=30mm,right=30mm,left=30mm,bottom=30mm]{geometry}
\usepackage[colorlinks=true,linkcolor=blue,urlcolor=blue]{hyperref}
\usepackage[affil-it,auth-sc]{authblk}
\usepackage[colorinlistoftodos,textwidth=25mm,textsize=footnotesize]{todonotes}
\usepackage{lipsum}                             
\usepackage[
	backend=biber,
	style=numeric,
	citestyle=numeric-comp,
	maxbibnames=99,
	minbibnames=99,
	maxcitenames=2,
	mincitenames=1,
	giveninits=true,
	sorting=none,
	sortlocale=auto,
	natbib=true,
	url=false,
	isbn=false,
	doi=true,
	eprint=true
]{biblatex}
\input{definitions}                             
\graphicspath{{./}{./figures/}}                 

\title{Tensile material instabilities in elastic beam lattices lead to a bounded stability domain}

\author[1]{G. Bordiga}
\author[1]{D. Bigoni\footnote{Corresponding author: e-mail: \href{mailto:bigoni@ing.unitn.it}{bigoni@ing.unitn.it}; phone: +39\,0461\,282507.}}
\author[1]{A. Piccolroaz}

\affil[1]{Department of Civil, Environmental, and Mechanical Engineering, University of Trento, Trento, Italy}

\date{\today}

\begin{document}

\maketitle

\begin{abstract}
	\noindent
	Homogenization of the incremental response of grids made up of \textit{preloaded} elastic rods leads to homogeneous effective continua which may suffer macroscopic instability, occurring at the same time in both the grid and the effective continuum.
	This instability corresponds to the loss of ellipticity in the effective material and the formation of localized responses as, for instance, shear bands.
	Using lattice models of elastic rods, loss of ellipticity has always been found to occur for stress states involving compression of the rods, as \textit{usually} these structural elements  buckle only under compression.
	In this way, the locus of material stability for the effective solid is unbounded in tension, i.e. the material is always stable for a tensile prestress.
	A rigorous application of homogenization theory is proposed to show that the inclusion of sliders (constraints imposing axial and rotational continuity, but allowing shear jumps) in the grid of rods leads to loss of ellipticity in tension, so that the locus for material instability becomes \textit{bounded}.
	This result explains (i.) how to design elastic materials subject to localization of deformation and shear banding for all radial stress paths; (ii.) how for all these paths a material may fail by developing strain localization and without involving cracking.
\end{abstract}

\paragraph{Keywords}
Tensile buckling \textperiodcentered\
Sliding interface \textperiodcentered\
Material instability \textperiodcentered\
Homogenization

\section{Introduction}
\label{sec:introduction}
A design strategy leading to metamaterials capable of effectively filtering and conditioning wave propagation is the composition of elastic structures via periodic lattices~\cite{craster_2012,torrent_2013,mcphedran_2015,piccolroaz_2020,nieves_2020,mishuris_2020,mishuris_2020a,nieves_2021,madine_2021}.
In these structures, different effects related to out-of-plane or in-plane deformations, presence of bending moment or prestress have been explored~\cite{carta_2017,garau_2019a,bordiga_2019a,bordiga_2019}.
Still, many important issues remain unknown, so that the present article addresses one of these, namely, the possibility of defining structured materials capable of suffering instabilities for all possible prestress states, including \textit{tensile}.

The incremental response of a periodic grid of preloaded elastic rods can be homogenized to obtain an effective, prestressed elastic solid, linearly relating the increments of the first Piola-Kirchhoff stress and of the displacement gradient,~\cite{triantafyllidis_1985,triantafyllidis_1993,nestorovic_2004,santisidavila_2016}.
Two types of instability may occur in the grid, classified as `microscopic' and `macroscopic'.
Only the latter is captured by the homogenized material response and corresponds to its loss of ellipticity, which, in turn, coincides with the condition of strain localization, and, as a special case, shear band formation~\cite{pontecastaneda_1989,pontecastaneda_1996,pontecastaneda_1997,avazmohammadi_2016,furer_2018}.

Recently, two-dimensional grids of prestressed elastic rods, subject to in-plane incremental normal and shear forces and bending moment, have been advocated as materials that can be designed to exhibit instabilities inside the elastic range as well as to display tunable effective properties~\cite{bordiga_2021}.
However, loss of ellipticity in these materials has been so far shown to be possible only when compression is involved, so that the locus of material stability for the homogenized material is \textit{unbounded in tension}.
This circumstance is a direct consequence of the fact that \textit{usually} elastic rods only buckle in compression.
However, real materials exhibit localization of strain for all stress states, including tension~\cite{bigoni_2012}.
Therefore, it might be superficially (and erroneously) concluded that it is impossible to design an artificial material with a \textit{bounded} stability domain, using a grid of elastic rods.
In contrast with this erroneous conclusion, it is shown in this article that employing slider constraints (permitting only relative transverse sliding between the connected ends of two rods) inside the rods forming the periodic lattice may lead to a macroscopic buckling in tension, formally corresponding to loss of ellipticity in the homogenized response for tensile stress states.

As a consequence, it is rigorously proven that homogenization leads to a \textit{smooth and bounded domain where failure determined by strain localization and shear bands is excluded}.
This failure occurs as soon as this domain is touched, as it happens for every radial stress paths emanating from the unloaded state.

\begin{figure}[htbp]
	\centering
	\includegraphics[width=0.98\linewidth]{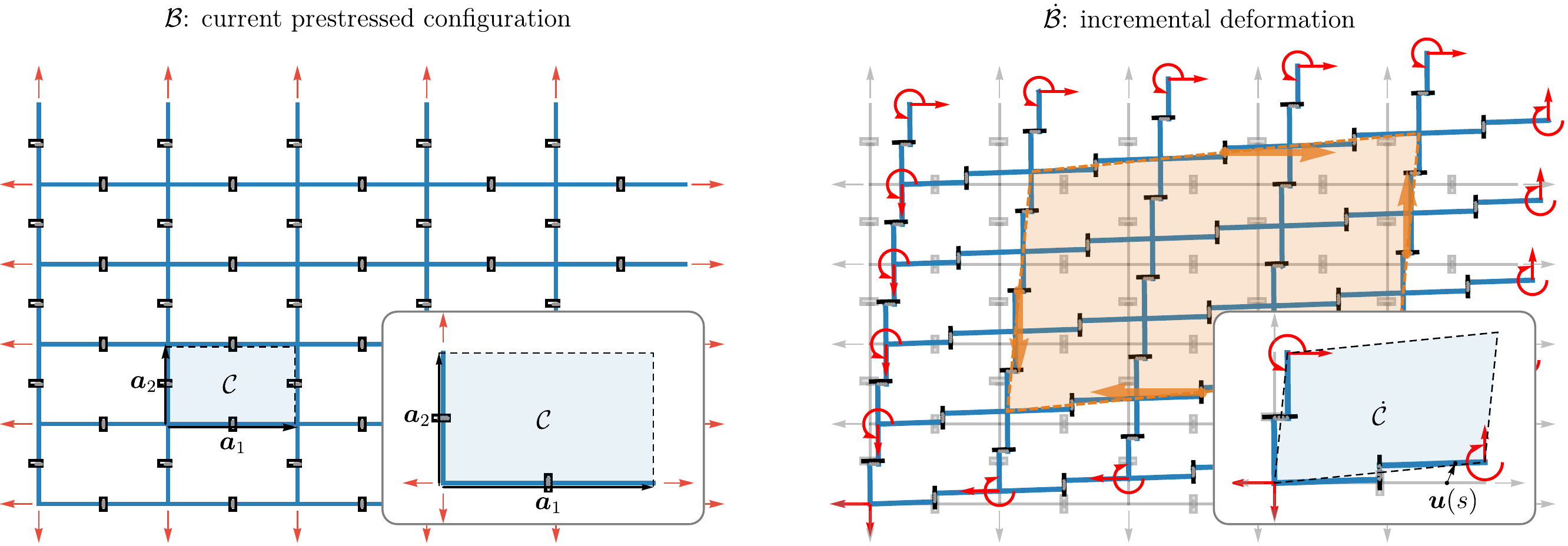}
	\caption{
		A periodic two-dimensional grid of (axially and flexurally deformable) elastic rods equipped with sliders leads to \textit{tensile and compressive} global bifurcations corresponding to loss of strong ellipticity in the effective material.
		The rods are axially preloaded in tension or compression from a stress-free configuration (left).
		The response to an incremental shear displacement $\bu(s)$ on the boundary of the grid leads to shear forces and bending moments (shown in red) in the rods and it corresponds in the effective material to an incremental shear (shown in orange), so that the bending moments in the rods do not contribute to the mean stress (right).
	}
	\label{fig:lattice_configurations}
\end{figure}

Therefore, a new way is found to design architected materials with a stability (or `failure') domain that is bounded for all stress `directions'.
These materials, designed as grids of elastic beams endowed with slider constraints as sketched in Fig.~\ref{fig:lattice_configurations}, will be shown to greatly extend their compliance under stretching as a consequence of the occurrence of a localized shear deformation band.

The methodology developed in this article allows for the control and design of the bifurcation pattern triggered by both compressive and tensile stress in a 2D lattice material.
These can be effectively leveraged for a wide range of engineering applications, as recent studies show that suitably controlled buckling instabilities can be harnessed to realize innovative devices.
Some instances of direct exploitation of a specific kind of `tensile' buckling (namely an instability triggered by local compression produced by a global stretch) have been explored for the design of flexible 3D electronics~\cite{guo_2018,cheng_2019} and bio-inspired active skins~\cite{park_2019}.
Therefore, the extension of the design space to structures and materials exhibiting buckling under \textit{pure} tension can open unexplored routes to novel fabrication processes for periodic electro-mechanical systems at different scales.

The results shown in the present article are directly connected to the discovery of tensile buckling~\cite{zaccaria_2011,bigoni_2018} and are obtained through a rigorous application of homogenization theory providing an energetic match between a preloaded lattice (Section~\ref{sec:prestressed_lattice}) and an effective elastic continuum (Section~\ref{sec:homogenization}).
Examples of materials characterized by a bounded stability domain and their characteristics are provided (Section~\ref{sec:grid}), followed by the analysis of the `re-stabilization' occurring in the effective continuum, while the elastic grid is subject to local instabilities (Section \ref{sec:restab}). 
Finally, concluding remarks are drawn at the end of the paper (Section \ref{sec:concluding}).

\section{Incremental response of lattices incorporating sliding constraints}
\label{sec:prestressed_lattice}
A two-dimensional periodic lattice of elastic rods, deformable in the plane both axially and flexurally, is considered, in which all structural members are axially prestressed from an unloaded reference configuration $\mB_0$.
Each junction between the rods is assumed to be one of two types: a fully welded junction or a sliding constraint.
The former is standard as it prescribes perfect continuity of displacements and rotations, while the latter allows for a displacement jump along the sliding direction, here assumed to be orthogonal to the rod's axis, see Fig.~\ref{fig:lattice_configurations}.

The prestress is produced by tensile or compressive dead loading acting at infinity, while body forces in the lattice are not considered for simplicity.
The preload is postulated not only to satisfy equilibrium, but also to preserve periodicity and leave the structure free from flexure.
The incremental response is analyzed by considering arbitrary deformations, which include development of bending moment and axial and shear forces.

The prestressed configuration $\mB$ is periodic along two linearly independent vectors $\{\ba_1,\ba_2\}$, defining the direct basis of the lattice, so that the structure can be constructed from a single unit cell $\mC$, assumed to be composed of $N_{\text{B}}$ nonlinear elastic rods with Euler-Bernoulli incremental kinematics, as sketched in Fig.~\ref{fig:lattice_configurations}.

By considering in-plane flexural and axial incremental deformations, the incremental displacement field of the $k$--th rod in a given unit cell is defined by the vector field
\begin{equation}
	\label{eq:displacement_beams}
	\bu_k(s_k)=\trans{\{ u_k(s_k), v_k(s_k)\}}, \quad \forall k \in \{1,...,N_{\text{B}}\} \,,
\end{equation}
where $s_k$ is the coordinate along the $k$--th rod, $u_k(s_k)$ and $v_k(s_k)$ are the axial and transverse incremental displacements.
The incremental rotation of the rod's cross-section $\theta_k(s_k)$ is assumed to satisfy the unshearability condition of the Euler-Bernoulli kinematics, namely, $\theta_k(s_k)=v_k'(s_k)$.

In order to formulate the problem of incremental equilibrium for the lattice, the contributions to the second-order incremental energy are derived for the single rod in Section~\ref{sec:solution_beam} and for the slider constraint in Section~\ref{sec:slider_contribution}.
These are then combined in Section~\ref{sec:equilibrium_lattice} to obtain the unit cell equilibrium.

\subsection{Analytic solution for the prestressed elastic rod}
\label{sec:solution_beam}
The incremental equilibrium equations for an elastic rod obeying Euler-Bernoulli kinematics, prestressed with an axial load $P$ (assumed positive in tension), and pre-stretched by $\lambda_0>0$, are the following
\begin{subequations}
	\label{eq:governing_beam_EB}
	\begin{gather}
		\label{eq:governing_beam_EB_u}
		A(\lambda_0)\, u''(s) = 0 \,, \\
		\label{eq:governing_beam_EB_v}
		B(\lambda_0) \, v''''(s) - P(\lambda_0)\,v''(s) = 0 \,,
	\end{gather}
\end{subequations}
where $A(\lambda_0)$ and $B(\lambda_0)$ are the \textit{incremental} axial and bending stiffnesses, respectively, and $s\in(0,l)$ with $l$ being the \textit{current} length of the rod.
It is worth noting that the current axial and bending stiffnesses are, in general, functions of the current pre-stretch $\lambda_0$, which in turn depends on the axial load $P$ (see for instance~\cite{bordiga_2021}).
In the following, the parameters $A(\lambda_0)$ and $B(\lambda_0)$ will simply be denoted as $A$ and $B$, and treated as independent quantities for generality.

Eqs.~\eqref{eq:governing_beam_EB} is a system of linear ODEs for the functions $u(s)$ and $v(s)$.
As the system is fully decoupled, the solution is easily obtained in the form
\begin{equation}
	\label{eq:u_v_sol}
	u(s) = C_{1}^u + C_{2}^u\,s \,, \qquad v(s) = C_{1}^v\, e^{-\beta\,s} + C_{2}^v\, e^{\beta\,s} + C_{3}^v\,s + C_{4}^v \,,
\end{equation}
where $\{C_{1}^u,C_{2}^u,C_{1}^v,...,C_{4}^v\}$ are 6 arbitrary complex constants and $\beta=\sqrt{P/B}$.

For a rod of length $l$, the following nomenclature can be introduced
\begin{equation}
	\label{eq:beam_bcs}
	u(0) = u_1 \,, \quad v(0) = v_1 \,, \quad \theta(0) = \theta_1 \,, \quad u(l) = u_2 \,, \quad v(l) = v_2 \,, \quad \theta(l) = \theta_2 \,,
\end{equation}
so that the vector $\bq=\trans{\{u_1,v_1,\theta_1,u_2,v_2,\theta_2\}}$ collects the degrees of freedom of the rod expressed in terms of end displacements.
Solving the conditions~\eqref{eq:beam_bcs} for the constants $\trans{\{C_{1}^u,C_{2}^u,C_{1}^v,...,C_{4}^v\}}$ allows the solution~\eqref{eq:u_v_sol} to be rewritten as
\begin{equation}
	\label{eq:static_sf}
	\bu(s) = \bN(s;P)\, \bq \,,
\end{equation}
which is now a linear function of the nodal displacements $\bq$.
The $2{\times}6$ matrix $\bN(s;P)$ acts as a matrix of prestress-dependent `shape functions' and therefore the representation~\eqref{eq:static_sf} can also be considered as the definition of a `finite element' endowed with shape functions built from the exact solution.
Moreover, these shape functions reduce to the solution holding true in the absence of prestress, because in the limit
\begin{equation*}
	\lim_{P\to0}\bN(s;P) =
	\begin{bmatrix}
		1-\frac{s}{l} & 0                           & 0                     & \frac{s}{l} & 0                         & 0                     \\[1mm]
		0             & \frac{(l-s)^2 (l+2 s)}{l^3} & \frac{(l-s)^2 s}{l^2} & 0           & \frac{(3 l-2 s) s^2}{l^3} & \frac{s^2 (s-l)}{l^2}
	\end{bmatrix} \,,
\end{equation*}
the usual shape functions (linear and Hermitian for axial and flexural displacements, respectively) are retrieved.

By employing Eq.~\eqref{eq:static_sf}, the incremental stiffness matrix of a prestressed rod can be computed, so that
for the $k$-th rod the elastic strain energy at second order is given by
\begin{equation}
	\label{eq:strain_energy_beam}
	\mE_k = \frac{1}{2} \int_0^{l_k} \left(A_k\,u'_k(s_k)^2 + B_k\,v''_k(s_k)^2\right) \,\text{d}s_k
	= \frac{1}{2} \,\trans{\bq_{k}} \left(\int_0^{l_k} \trans{\bB_{k}(s_k;P_k)}\bE_k\,\bB_{k}(s_k;P_k)\,\text{d}s_k \right) \bq_{k} \,,
\end{equation}
where $\bE_{k}$ is a matrix collecting the stiffness terms, while $\bB_{k}(s_k;P)$ is the strain-displacement matrix, defined as follows
\begin{equation*}
	\bE_{k}=\begin{bmatrix}
		A_k & 0   \\
		0   & B_k \\
	\end{bmatrix} \,,
	\qquad
	\bB_{k}(s_k;P_k) = \begin{bmatrix}
		\deriv{}{s_k} & 0                 \\
		0             & \deriv{^2}{s_k^2} \\
	\end{bmatrix} \bN_{k}(s_k;P_k) \,.
\end{equation*}
The `geometric' contribution due to the presence of the axial prestress is now included in the potential energy,
\begin{equation}
	\label{eq:prestress_energy_beam}
	\mV_k^g = \frac{1}{2} P_k \int_0^{l_k} v'_k(s_k)^2 \,\text{d}s_k
	= \frac{1}{2} \,\trans{\bq_{k}} \left(P_k \int_0^{l_k} \trans{\bb_{k}(s_k;P_k)} \bb_{k}(s_k;P_k) \,\text{d}s_k \right) \bq_{k} \,,
\end{equation}
where $\bb_{k}(s_k;P_k)=\begin{bmatrix}0 & \deriv{}{s_k}\end{bmatrix}\bN_{k}(s_k;P_k)$ is a vector collecting the derivatives of the shape functions describing the transverse displacement $v$.
A combination of Eqs.~\eqref{eq:strain_energy_beam} and~\eqref{eq:prestress_energy_beam}, yields the potential energy for the $k$-th rod in the form
\begin{equation}
	\label{eq:potential_energy_beam}
	\mV_k = \mE_k + \mV_k^g \,.
\end{equation}
Note that, as the equilibrium equations for the rods have been linearized around an axially preloaded configuration, the potential~\eqref{eq:potential_energy_beam} represents the incremental potential energy with respect to the current configuration.

From Eqs.~\eqref{eq:strain_energy_beam},~\eqref{eq:prestress_energy_beam} and~\eqref{eq:potential_energy_beam} the prestress-dependent stiffness matrix is defined as
\begin{equation*}
	\label{eq:K_beam}
	\bK_{k}(P_k) =
	\int_0^{l_k} \trans{\bB_{k}(s_k;P_k)}\bE_{k}\,\bB_{k}(s_k;P_k)\,\text{d}s_k +
	P_k \int_0^{l_k} \trans{\bb_{k}(s_k;P_k)} \bb_{k}(s_k;P_k) \,\text{d}s_k ,
\end{equation*}
so that
\begin{equation*}
	\bK_{k}=
	\begin{bmatrix}
		\frac{A_k}{l_k}  & 0                                   & 0                                  & -\frac{A_k}{l_k} & 0                                   & 0                                  \\[2mm]
		0                & \frac{12 B_k}{l_k^3}\varphi_1(p_k)  & \frac{6 B_k}{l_k^2}\varphi_2(p_k)  & 0                & -\frac{12 B_k}{l_k^3}\varphi_1(p_k) & \frac{6 B_k}{l_k^2}\varphi_2(p_k)  \\[2mm]
		0                & \frac{6 B_k}{l_k^2}\varphi_2(p_k)   & \frac{4 B_k}{l_k}\varphi_3(p_k)    & 0                & -\frac{6 B_k}{l_k^2}\varphi_2(p_k)  & \frac{2 B_k}{l_k}\varphi_4(p_k)    \\[2mm]
		-\frac{A_k}{l_k} & 0                                   & 0                                  & \frac{A_k}{l_k}  & 0                                   & 0                                  \\[2mm]
		0                & -\frac{12 B_k}{l_k^3}\varphi_1(p_k) & -\frac{6 B_k}{l_k^2}\varphi_2(p_k) & 0                & \frac{12 B_k}{l_k^3}\varphi_1(p_k)  & -\frac{6 B_k}{l_k^2}\varphi_2(p_k) \\[2mm]
		0                & \frac{6 B_k}{l_k^2}\varphi_2(p_k)   & \frac{2 B_k}{l_k}\varphi_4(p_k)    & 0                & -\frac{6 B_k}{l_k^2}\varphi_2(p_k)  & \frac{4 B_k}{l_k}\varphi_3(p_k)
	\end{bmatrix}
	,
\end{equation*}
where the $\varphi_j$ are functions of the non-dimensional measure of prestress $p_k=P_k l_k^2/B_k$ given by
\begin{align*}
	\varphi_1(p_k) & = \frac{p_k^{3/2}}{12 \left(\sqrt{p_k}-2 \tanh \left(\sqrt{p_k}/2\right)\right)} \,,                                                                                       &
	\varphi_2(p_k) & = \frac{p_k}{6 \sqrt{p_k} \coth \left(\sqrt{p_k}/2\right)-12} \,,                                                                                                            \\
	\varphi_3(p_k) & = \frac{p_k \cosh \left(\sqrt{p_k}\right)-\sqrt{p_k} \sinh \left(\sqrt{p_k}\right)}{4 \sqrt{p_k} \sinh \left(\sqrt{p_k}\right)-8 \cosh \left(\sqrt{p_k}\right)+8} \,,      &
	\varphi_4(p_k) & = \frac{\sqrt{p_k} \left(\sinh \left(\sqrt{p_k}\right)-\sqrt{p_k}\right)}{\left(4 \sqrt{p_k} \coth \left(\sqrt{p_k}/2\right)-8\right)\sinh^2\left(\sqrt{p_k}/2\right)} \,.
\end{align*}

Note that the tangent stiffness matrix $\bK_{k}$ representative of the $k-$rod in the prestressed lattice reduces, in the limit of vanishing prestress (or unitary pre-stretch $\lambda_{0k}=1$), to the usual stiffness matrix of an Euler-Bernoulli beam with Hermitian shape functions, so that
\begin{equation*}
	\lim_{p\to0} \varphi_j(p) = 1, \quad \forall j\in\{1,...,4\}.
\end{equation*}

\subsection{Second-order energy contribution of the slider constraint}
\label{sec:slider_contribution}
%
%
For the formulation of the incremental equilibrium of the lattice, the contributions to the potential energy of the  constraints between the rods need to be introduced.
As the constraints considered in this work are clamps and sliders, the derivation of the incremental contribution of the latter is addressed in the following, while the former simply impose continuity of displacements and rotations.

\begin{figure}[htbp]
	\centering
	\includegraphics[width=0.98\linewidth]{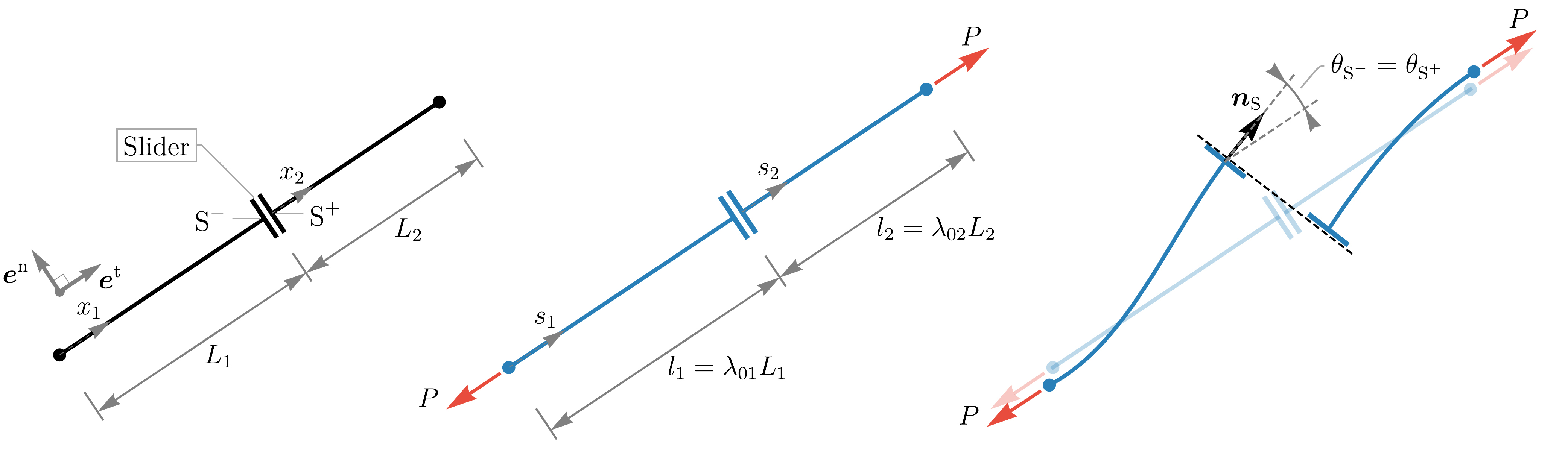}
	\caption{
	Stress-free (left), stretched (center), and incrementally  deformed (right) configurations of two rods connected to each other through a slider constraint and subject to an axial load $P$.
	The slider imposes a constraint on the displacement jump, $\jump{\bu}_{\text{S}}\scalp\bn_{\text{S}}=0$, and on the rotation, $\jump{\theta}_{\text{S}}=0$.
	}
	\label{fig:beams_slider}
\end{figure}

Two rods subject to the same axial load $P$ are considered, connected through a slider constraint, as sketched in Fig.~\ref{fig:beams_slider}.
By denoting the strain-energy density of the rods as $\psi$, the potential energy can be written as
\begin{equation}
	\label{eq:potential_energy_sliding_beams}
	\mV = \int_0^{L_1} \psi \, \text{d}x_1 + \int_0^{L_2} \psi\, \text{d}x_2 - P \, u_2(L_2) + P \, u_1(0) \,,
\end{equation}
where the local coordinates $x_1$ and $x_2$ as well as the rods' length are referred to the stress-free reference configuration.
The strain-energy density $\psi$ is dependent on the local stretch $\lambda$ and curvature $\chi$ of the rod, which are defined as
\begin{equation*}
	\label{eq:strain_measures_beam}
	\lambda = \left(1 + u'(x)\right) \cos\theta(x) + v'(x) \sin\theta(x) \,, \quad \chi = \theta'(x) \,,
\end{equation*}
with the rotation field $\theta$ satisfying the unshearability constraint $\theta=\arctan \left[v'/(1+u')\right]$.
The dependence of $\psi$ on $\lambda$ and $\chi$ is assumed in the form $\psi(\lambda,\chi)=\psi_{\lambda}(\lambda)+\psi_{\chi}(\chi)$.

The equilibrium configuration of the connected rods is provided by the principle of virtual work, effective to the stationary condition
\begin{equation}
	\label{eq:plv_sliding_beams}
	\delta\mV = \int_0^{L_1} \delta\psi\, \text{d}x_1 + \int_0^{L_2} \delta\psi\, \text{d}x_2 - P \,\delta u_2(L_2) + P \,\delta u_1(0) = 0 \,,
\end{equation}
whose solution is to be sought in the space of the displacement fields \textit{satisfying the slider constraint} between points $\text{S}^-$ and $\text{S}^+$, which reads
\begin{equation}
	\label{eq:slider_constraint}
	\jump{\bu}_{\text{S}} \scalp \underbrace{\left( \cos\theta_{\text{S}}\,\be^{\text{t}} + \sin\theta_{\text{S}}\,\be^{\text{n}} \right)}_{\bn_{\text{S}}} = 0 \,,
\end{equation}
where $\jump{\bu}_{\text{S}}$ denotes the displacement jump across the slider, $\jump{\bu}_{\text{S}} = \bu_{\text{S}^+} - \bu_{\text{S}^-}$, the unit vector $\bn_{\text{S}}$ describes the orientation of slider in the \textit{deformed} configuration, and the rotation field is continuous,  $\theta_{\text{S}}=\theta_{\text{S}^-}=\theta_{\text{S}^+}$, as sketched in Fig.~\ref{fig:beams_slider}.

Instead of dealing with Eq.~\eqref{eq:plv_sliding_beams} subject to the constraint~\eqref{eq:slider_constraint}, the equilibrium can be equivalently formulated by means of the Lagrange multiplier method, so that the potential energy~\eqref{eq:potential_energy_sliding_beams} becomes
\begin{equation}
	\label{eq:potential_energy_sliding_beams_constrained}
	\hat{\mV} = \underbrace{\int_0^{L_1} \psi \, \text{d}x_1 + \int_0^{L_2} \psi\, \text{d}x_2 - P \, u_2(L_2) + P \, u_1(0)}_{\mV} + \underbrace{\mu\,\jump{\bu}_{\text{S}} \scalp \left( \cos\theta_{\text{S}}\,\be^{\text{t}} + \sin\theta_{\text{S}}\,\be^{\text{n}} \right)}_{\mS} ,
\end{equation}
with $\mu$ being the Lagrange multiplier of the slider constraint.
The augmented potential~\eqref{eq:potential_energy_sliding_beams_constrained} is now used to obtain the straight stretched equilibrium configuration and the second-order potential governing the incremental equilibrium.

The straight equilibrium configuration is easily obtained by solving the vanishing condition of the first variation of functional~\eqref{eq:potential_energy_sliding_beams_constrained},  evaluated for a displacement field of the form
\begin{equation*}
	u_{0k}(x_k) = u_{0k}(0) + (u_{0k}(L_k)-u_{0k}(0))\frac{x_k}{L_k} \,, \qquad v_{0k}(x_k) = 0 \, \qquad \forall k\in\{1,2\} \,,
\end{equation*}
so that stationarity of the augmented potential \eqref{eq:potential_energy_sliding_beams_constrained} reads as
\begin{equation}
	\label{eq:first_variation_potential_energy_sliding_beams_constrained}
	\begin{multlined}[0.9\linewidth]
		\delta\hat{\mV}_0 = \int_0^{L_1} \psi_{\lambda}'(\lambda_{01})\,\delta u_1' \, \text{d}x_1 + \int_0^{L_2} \psi_{\lambda}'(\lambda_{02})\,\delta u_2' \, \text{d}x_2 - P \, \delta u_2(L_2) + P \, \delta u_1(0) + \\
		+ \mu \left( \delta u_2(0) - \delta u_1(L_1) \right) + \delta \mu \left( u_{02}(0) - u_{01}(L_1) \right) = 0 \,, \quad \forall \delta u_1, \delta u_2, \delta \mu \,,
	\end{multlined}
\end{equation}
yielding the following equilibrium conditions
\begin{equation}
	\label{eq:straight_equilibrium_sliding_beams}
	\psi_{\lambda}'(\lambda_{01}) = \psi_{\lambda}'(\lambda_{02}) = \mu_0 = P \,, \qquad u_{02}(0) = u_{01}(L_1) \,,
\end{equation}
where $\lambda_{01}$ and $\lambda_{02}$ denote the stretch values of the rods, while $\mu=\mu_0$ denotes the value of the Lagrange multiplier at the equilibrium.
These conditions are derived from equation~\eqref{eq:first_variation_potential_energy_sliding_beams_constrained} by taking advantage of homogeneity of the stretch in the two rods (so that $\lambda_{01}$ and $\lambda_{02}$ are constants) and noting also that the residual bending moment $\psi_{\chi}'(0)$ vanishes on a straight configuration.

Upon the straight stretched configuration defined by Eq.~\eqref{eq:straight_equilibrium_sliding_beams}, the \textit{incremental} equilibrium is governed by the \textit{second-order} expansion of the augmented potential, Eq.~\eqref{eq:potential_energy_sliding_beams_constrained}.
Letting $\Delta\bu_1$, $\Delta\bu_2$, and $\Delta\mu$ be the increments with respect to the straight equilibrium configuration, the expansion assumes the form
\begin{equation}
	\label{eq:second_order_expansion_potential_sliding_beams}
	\hat{\mV}(\bu_{01}+\Delta\bu_1,\bu_{02}+\Delta\bu_2,\mu_0+\Delta\mu) \sim \underbrace{\delta\hat{\mV}_0}_{=0} + \delta^2\mV_0(\Delta\bu_1,\Delta\bu_2,\Delta\mu) + \delta^2\mS_0(\Delta\bu_1,\Delta\bu_2,\Delta\mu) ,
\end{equation}
where the subscript $(\cdot)_0$ highlights reference to the straight equilibrium configuration.

The second-order part of Eq.~\eqref{eq:second_order_expansion_potential_sliding_beams} is composed of two terms: the term $\delta^2\mV_0$ accounting for the incremental energy distributed on the rods (as if the slider were absent) and the term $\delta^2\mS_0$ that accounts only for the contribution of the slider constraint.
The rods' contribution $\delta^2\mV_0$ assumes the usual form, Eq.~\eqref{eq:potential_energy_beam} (see~\cite{bordiga_2021} for details), while the slider contribution $\delta^2\mS_0$ has to be determined.
Using the definition of $\mS$ from Eq.~\eqref{eq:potential_energy_sliding_beams_constrained} and recalling the properties of the straight equilibrium configuration, Eqs.~\eqref{eq:straight_equilibrium_sliding_beams}, $\delta^2\mS_0$ becomes
\begin{equation}
	\label{eq:second_order_potential_slider_constraint}
	\begin{aligned}
		\delta^2\mS_0(\Delta\bu_1,\Delta\bu_2,\Delta\mu)
		 & = \mu_0\,\jump{\Delta\bu}_{\text{S}} \scalp \Delta\theta_{\text{S}}\be^{\text{n}} + \Delta\mu\,\jump{\Delta\bu}_{\text{S}} \scalp \be^{\text{t}} \\
		 & = \mu_0(\Delta v_2(0)-\Delta v_1(L_1))\Delta\theta_{\text{S}} + \Delta\mu(\Delta u_2(0)-\Delta u_1(L_1)) \,,
	\end{aligned}
\end{equation}
where the incremental rotation of the slider $\Delta\theta_{\text{S}}$ can be written in terms of the transverse displacement as
\begin{equation*}
	\Delta\theta_{\text{S}} = \frac{1}{\lambda_{01}}\deriv{\Delta v_1}{x_1}(L_1)	= \frac{1}{\lambda_{02}}\deriv{\Delta v_2}{x_2}(0) \,,
\end{equation*}
or also as
\begin{equation*}
	\Delta\theta_{\text{S}} = \deriv{\Delta v_1}{s_1}(l_1) = \deriv{\Delta v_2}{s_2}(0) \,,
\end{equation*}
obtained by simply updating the reference configuration from the stress-free, described by the coordinate $x_k\in[0,L_k]$, to the current stretched configuration function of  $s_k=\lambda_{0k}x_k\in[0,l_k]$.

Having determined the second-order contribution of the slider, Eq.~\eqref{eq:second_order_potential_slider_constraint}, the symbol $\Delta$, introduced to denote incremental quantities, will be omitted in order to ease the notation.
Thus, all quantities are assumed in the following to be incremental quantities unless stated otherwise.
Accordingly, the contribution of the slider, Eq.~\eqref{eq:second_order_potential_slider_constraint}, is simply  denoted by $\mS$ and written as
\begin{equation}
	\label{eq:slider_contribution}
	\mS\left(\jump{\bu}_{\text{S}},\mu\right) = P \,\jump{\bu}_{\text{S}} \scalp \theta_{\text{S}}\be^{\text{n}} + \mu\,\jump{\bu}_{\text{S}} \scalp \be^{\text{t}} \,,
\end{equation}
where the equality $\mu_0=P$ provided by  Eq.~\eqref{eq:straight_equilibrium_sliding_beams} has been used.
The stationary of Eq.~\eqref{eq:slider_contribution} with respect to the Lagrange multiplier $\mu$ yields the condition $\jump{\bu}_{\text{S}} \scalp \be^{\text{t}}=0$, expressing continuity of the axial displacement across the slider.
Hence, Eq.~\eqref{eq:slider_contribution} is finally simplified as
\begin{equation}
	\label{eq:slider_contribution_simplified}
	\mS\left(\jump{\bu}_{\text{S}}\right) = P \,\jump{\bu}_{\text{S}} \scalp \theta_{\text{S}}\be^{\text{n}} \,.
\end{equation}

The following subsection combines the incremental energy of the rods, Eq.~\eqref{eq:potential_energy_beam}, with that pertaining to the sliders, Eq.~\eqref{eq:slider_contribution_simplified}, to construct the incremental equilibrium of the entire lattice structure.

\subsection{Incremental equilibrium for the unit cell}
\label{sec:equilibrium_lattice}
The equations governing the incremental equilibrium are formulated for a unit cell of the lattice with respect to the current preloaded configuration.
The incremental potential energy $\mV(\bq)$ of a unit cell can be evaluated by summing the contribution of each rod, Eq.~\eqref{eq:potential_energy_beam}, as well as of each slider, Eq.~\eqref{eq:slider_contribution_simplified}, so that
\begin{equation}
	\label{eq:potential_energy_unit_cell}
	\mV(\bq) = \sum_{k=1}^{N_{\text{B}}} \mV_k(\bC_k \bq) + \sum_{i=1}^{N_{\text{S}}} \mS_i(\jump{\bu}_{S_i}) \,,
\end{equation}
where $N_{\text{S}}$ is the number of sliders in the unit cell, $\bq$ is the vector collecting the degrees of freedom of the unit cell, $\mS_i$ denotes the contribution of $i$-th slider, and $\bC_k$ is the connectivity matrix of the $k$-th rod, such that $\bq_k = \bC_k \bq$, which imposes the appropriate constraints at the junctions between the rods (continuity of all the fields for welded junctions and continuity of rotation and axial displacement for sliders).

The current configuration of the unit cell is subject to \textit{external generalized incremental} forces $\bef$ (including bending moments) transmitted by the rest of the lattice from which the cell is thought to be ideally `excised'.
Hence, the incremental equilibrium for a single unit cell can be stated through the principle of virtual work as
\begin{equation}
	\label{eq:plv_unit_cell}
	\delta\mV(\bq,\delta\bq) = \bef\scalp\delta\bq\,, \quad \forall\delta\bq \,,
\end{equation}
where $\delta\bq$ is the virtual counterpart of $\bq$.
Note that, due to the assumption of absence of body forces, the \textit{external} virtual work $\bef\scalp\delta\bq$ only involves forces applied on the unit cell boundary, as the only non-vanishing external forces acting on a unit cell are those transmitted by the neighboring cells.

The incremental equilibrium equations are therefore obtained from the variational statement, Eq.~\eqref{eq:plv_unit_cell}, yielding
\begin{equation}
	\label{eq:equilibrium_unit_cell}
	\bK(\bP)\,\bq = \bef \,,
\end{equation}
where
\begin{equation}
	\bK(\bP) = \frac{\partial^2 \mV(\bq)}{\partial\bq\,\partial\bq} \,,
\end{equation}
is the symmetric (as derived from a scalar potential) stiffness matrix of the unit cell, function of the vector $\bP=\{P_1,...,P_{N_{\text{B}}}\}$, which collects the axial prestress of the rods.
The dimension of the system~\eqref{eq:equilibrium_unit_cell} is $3N_j$ where $N_j$ is the number of nodes in the unit cell.

It is worth pointing out that the contribution $\mV_k$ from each rod,  appearing in the potential energy, Eq.~\eqref{eq:potential_energy_unit_cell}, is positive definite whenever the preload is tensile ($P_k>0$), while the slider contribution $\mS_k$ is \textit{indefinite} even for a tensile preload.
This property implies that in the absence of slider constraints bifurcations would be excluded for tensile preload state.
Therefore, the presence of the sliders allows $\mV$ to vanish for a non-trivial deformation even when $P_k>0\,,\,\forall k \in \{1,...,N_{\text{B}}\}$ and thus tensile instabilities of the lattice (and its effective continuum) become possible.

\section{The effective prestressed elastic continuum}
\label{sec:homogenization}
As the preloaded configuration of the lattice is assumed to be spatially periodic, the homogenized incremental response of an effective prestressed elastic solid can be defined by computing the average strain-energy density, associated to an incremental displacement field (defined for the $j$-th node by the displacement and rotation components, respectively, $\bq_u^{(j)} = \trans{\{u^{(j)},v^{(j)}\}}$ and $\bq_\theta^{(j)} = \{\theta^{(j)}\}$) which obeys the Cauchy-Born hypothesis~\cite{born_1955,willis_2002,hutchinson_2006}.
The latter, for a single unit cell, prescribes that the displacement of the lattice's nodes be decomposed into the sum of an affine incremental deformation (ruled by a constant second-order tensor $\bL$) and a periodic field (defined by a displacement $\Tilde{\bq}_u^{(j)}$ and a rotational $\Tilde{\bq}_\theta^{(j)}$ component) as
\begin{equation}
	\label{eq:cauchy_born_hypothesis}
	\bq_u^{(j)} = \Tilde{\bq}_u^{(j)} + \bL\,\bx_j \,, \qquad \bq_\theta^{(j)} = \Tilde{\bq}_\theta^{(j)} \,, \qquad \forall j\in\{1,...,N_j\} \,,
\end{equation}
where $\bx_j$ is the position of the $j$-th node.

The periodic term $\Tilde{\bq}$ satisfies $\Tilde{\bq}^{(p)}=\Tilde{\bq}^{(q)}$ for all $\{p,q\}$ such that $\bx_q-\bx_p = n_1\ba_1+n_2\ba_2$ (with $n_j\in\{0,1\}$).
This term  can be expressed as a function of its independent components through a partition of the degrees of freedom, to be made in accordance with the location of the nodes present inside the unit cell.
Specifically, by denoting with $\Tilde{\bq}^i$ the degrees of freedom located inside the unit cell, with $\Tilde{\bq}^l$, $\Tilde{\bq}^r$, $\Tilde{\bq}^b$, $\Tilde{\bq}^t$ those on the left, right, lower, and upper edge respectively, and with $\Tilde{\bq}^{lb}$, $\Tilde{\bq}^{rb}$, $\Tilde{\bq}^{lt}$, $\Tilde{\bq}^{rt}$ those located at the four corners, the periodic field can be expressed as
\begin{subequations}
	\label{eq:periodic_conditions}
	\begin{equation}
		\label{eq:periodic_conditions_long}
		\Tilde{\bq} =
		\begin{Bmatrix}
			\Tilde{\bq}^i    \\
			\Tilde{\bq}^l    \\
			\Tilde{\bq}^b    \\
			\Tilde{\bq}^{lb} \\
			\Tilde{\bq}^r    \\
			\Tilde{\bq}^t    \\
			\Tilde{\bq}^{rb} \\
			\Tilde{\bq}^{lt} \\
			\Tilde{\bq}^{rt}
		\end{Bmatrix}
		=
		\begin{bmatrix}
			\bI    & \bzero & \bzero & \bzero \\
			\bzero & \bI    & \bzero & \bzero \\
			\bzero & \bzero & \bI    & \bzero \\
			\bzero & \bzero & \bzero & \bI    \\
			\bzero & \bI    & \bzero & \bzero \\
			\bzero & \bzero & \bI    & \bzero \\
			\bzero & \bzero & \bzero & \bI    \\
			\bzero & \bzero & \bzero & \bI    \\
			\bzero & \bzero & \bzero & \bI
		\end{bmatrix}
		\begin{Bmatrix}
			\Tilde{\bq}^i \\
			\Tilde{\bq}^l \\
			\Tilde{\bq}^b \\
			\Tilde{\bq}^{lb}
		\end{Bmatrix} \,,
	\end{equation}
	which may succinctly be rewritten as
	\begin{equation}
		\label{eq:periodic_conditions_short}
		\Tilde{\bq} = \bZ_0\,\Tilde{\bq}^* \,,
	\end{equation}
\end{subequations}
where $\bZ_0$ and $\Tilde{\bq}^*$ are defined according to Eq.~\eqref{eq:periodic_conditions_long}.
The same partitioning is also used for the vectors $\bq$ and $\bef$.
Note that the periodicity conditions~\eqref{eq:periodic_conditions} represent the long-wavelength limit of the Floquet-Bloch conditions used in wave propagation problems~\cite{phani_2006,bordiga_2021}.

In order to enforce the Cauchy-Born conditions into the equations of incremental equilibrium~\eqref{eq:equilibrium_unit_cell}, it is convenient to rewrite Eq.~\eqref{eq:cauchy_born_hypothesis} as
\begin{equation}
	\label{eq:cauchy_born_hypothesis_vector}
	\bq(\Tilde{\bq}^*,\bL) = \bZ_0\,\Tilde{\bq}^* + \hat{\bq}(\bL) \,,
\end{equation}
where the affine part of the deformation $\hat{\bq}(\bL)$ is a vector-valued function linear in $\bL$ and such that
\begin{equation*}
	\hat{\bq}(\bL)_u^{(j)}=\bL\,\bx_j \,, \qquad \hat{\bq}(\bL)_\theta^{(j)}=\bzero  \,, \qquad \forall j\in\{1,...,N_j\} \,,
\end{equation*}
where the same notation introduced with Eq. \eqref{eq:cauchy_born_hypothesis} has been used so that the subscript $u$ (subscript $\theta$) denotes displacement (rotation) components.

Note that, since the lattice is subject to a non-vanishing prestress state, the macroscopic incremental deformation gradient defined by $\bL$ must be an arbitrary second-order tensor, \textit{not constrained to be symmetric} (as it happens in the absence of prestress~\cite{pontecastaneda_1997,willis_2002,hutchinson_2006}).
As explained in the next section, this lack of symmetry is essential for the correct evaluation of the incremental fourth-order tensor defining the effective continuum, `macroscopically equivalent' to the lattice.

\subsection{Incremental constitutive tensor for the effective continuum}
\label{sec:homogenization_constitutive_tensor}
Before introducing the homogenization technique, it is important to recall that, as shown in Section~\ref{sec:prestressed_lattice}, the equilibrium equations for the lattice are
\begin{enumerate*}[label=(\roman*)]
	\item obtained in the context of a linearized theory, and
	\item referred to a prestressed reference configuration.
\end{enumerate*}
Therefore, the effective continuum, for the moment unknown, has to be formulated in the context of the incremental theory of nonlinear elasticity by means of a relative Lagrangian description as introduced by Hill~\cite{hill_1957}, see also~\cite{bigoni_2012}.
As a consequence, the response of the effective material is defined by an \textit{incremental constitutive law} in the form
\begin{equation}
	\label{eq:incremental_constitutive_law}
	\dot{\bS} = \fC[\bL] \,,
\end{equation}
relating the increment of the first Piola-Kirchhoff stress $\dot{\bS}$ to the gradient of incremental displacement $\bL$, through the elasticity tensor $\fC$.
The most general form for the constitutive tensor $\fC$ is
\begin{equation}
	\label{eq:constitutive_operator}
	\fC = \fE + \bI \boxtimes \bT \qquad \mbox{in components} \qquad \fC_{ijkl} = \fE_{ijkl} + \delta_{ik} T_{jl} \,,
\end{equation}
where $\delta_{ik}$ is the Kronecker delta, $\bT$ is the Cauchy stress, defining the \textit{prestress}, and $\fE$ is a fourth-order elastic tensor, endowed with all usual (left and right minor and major) symmetries
\begin{equation}
	\label{eq:minor_symmetries_E}
	\fE_{ijkl} = \fE_{jikl} = \fE_{ijlk} = \fE_{klij} \,,
\end{equation}
so that $\fC$ lacks the minor symmetries but possesses the major symmetry.
The symmetries of $\fC$ explain the reason why the full incremental deformation gradient $\bL$, and \textit{not} only its symmetric part, appears in the Cauchy-Born hypothesis, Eq.~\eqref{eq:cauchy_born_hypothesis}, of the lattice.
Moreover, Eq.~\eqref{eq:constitutive_operator} shows that $\bL$ \textit{can be restricted to be symmetric only in the absence of prestress}, $\bT=\bzero$.

The incremental strain-energy density for the prestressed continuum is referred to the prestressed configuration.
It can be expressed in terms of a second-order expansion with respect to the incremental deformation gradient $\bL$ as follows
\begin{equation}
	\label{eq:incremental_energy_continuum}
	\mW(\bL) = \underbrace{\bT\scalp\bL}_{\mW_1(\bL)} + \underbrace{\fC[\bL]\scalp\bL/2}_{\mW_2(\bL)} \,,
\end{equation}
where the first-order increment $\mW_1(\bL)$ accounts for the work expended by the current prestress state $\bT$ (due to the relative Lagrangian description the first Piola-Kirchhoff stress coincides with the Cauchy stress), while the second-order term $\mW_2(\bL)$ is the strain-energy density associated with the incremental first Piola-Kirchhoff stress given by Eq.~\eqref{eq:incremental_constitutive_law}.

It is also worth noting that a calculation of the second gradient of the incremental energy density, Eq.~\eqref{eq:incremental_energy_continuum}, with respect to $\bL$ yields the constitutive fourth-order tensor $\fC$ relating the stress increment to the incremental displacement gradient.
Taking the first gradient provides, when evaluated at $\bL=\bzero$, the prestress $\bT$.
The latter property will be used to dissect the effect of prestress in the homogenized response of the lattice.

\subsection{First and second-order matching of the incremental strain-energy density}
\label{sec:energy_matching}
The homogenization of the lattice response is based on the equivalence between the average incremental strain-energy associated to a \textit{macroscopic} incremental displacement gradient applied to the lattice and the incremental strain-energy density of the effective elastic material subject to the same deformation.
In the classical homogenization theory, this condition is known as \textit{macro-homogeneity} condition, or Hill-Mandel theorem~\cite{hill_1972,sanchez-palencia_1987,pontecastaneda_1997,willis_2002}, which provides the link between the microscopic and macroscopic scales.

In the following, the macro-homogeneity condition is enforced to obtain the incremental energy density~\eqref{eq:incremental_energy_continuum} that matches the effective behavior of the prestressed lattice at first- $\mW_1(\bL)$ and at second- $\mW_2(\bL)$ order.
Thus, the homogenization scheme is based on the following steps:
\begin{enumerate}[label=(\roman*)]
	\item An incremental deformation gradient $\bL$ is considered, so that the incremental energy density for the unknown effective continuum is defined by Eq.~\eqref{eq:incremental_energy_continuum};
	\item following the Cauchy-Born hypothesis,  Eq.~\eqref{eq:cauchy_born_hypothesis_vector}, the incremental displacement field for the lattice is prescribed by the given tensor $\bL$ and the periodic vector $\Tilde{\bq}^*$ necessary to enforce the  equilibrium of the lattice;
	\item with the solution of the lattice in terms of $\bL$ (the periodic vector $\Tilde{\bq}^*$ becomes in solution a function of $\bL$) the incremental energy density is calculated for the lattice;
	\item the two incremental energy densities in the continuum and in the lattice are matched, so to obtain the components of the incremental elastic tensor defining the effective solid.
\end{enumerate}

\paragraph{Determination of the periodic displacement field for the lattice.}
By substituting condition~\eqref{eq:cauchy_born_hypothesis_vector} into Eqs.~\eqref{eq:equilibrium_unit_cell} and pre-multiplying by $\trans{\bZ_0}$, the incremental equilibrium becomes
\begin{equation}
	\label{eq:equilibrium_internal_strain_f}
	\trans{\bZ_0}\bK(\bP)\,\bZ_0\,\Tilde{\bq}^* + \trans{\bZ_0}\bK(\bP)\,\hat{\bq}(\bL) = \trans{\bZ_0}\bef \,,
\end{equation}
where the right-hand side can be written more explicitly using the partitioning introduced by Eq.~\eqref{eq:periodic_conditions_long} as
\begin{equation*}
	\trans{\bZ_0}\bef =
	\begin{Bmatrix}
		\bef^i          \\
		\bef^l + \bef^r \\
		\bef^b + \bef^t \\
		\bef^{lb} + \bef^{rb} + \bef^{lt} + \bef^{rt}
	\end{Bmatrix} \,.
\end{equation*}
The fact that the only non-vanishing forces are assumed to be the internal forces transmitted at the unit cell boundary by the neighboring cells implies $\bef^i=\bzero$.
Moreover, as the displacement field satisfying the Cauchy-Born hypothesis generates \textit{internal} forces in the infinite lattice that are \textit{periodic} along the direct basis $\{\ba_1,\ba_2\}$, any single unit cell is subject to \textit{external} boundary forces that are \textit{anti-periodic}.
Consequently, $\bef^l=-\bef^r$, $\bef^b=-\bef^t$ and $\bef^{lb}=-\bef^{rb}-\bef^{lt}-\bef^{rt}$, so that the term $\trans{\bZ_0}\bef$ vanishes and Eq.~\eqref{eq:equilibrium_internal_strain_f} becomes
\begin{equation}
	\label{eq:equilibrium_internal_strain}
	\trans{\bZ_0}\bK(\bP)\,\bZ_0\,\Tilde{\bq}^* = -\trans{\bZ_0}\bK(\bP)\,\hat{\bq}(\bL) \,.
\end{equation}
The solution of the linear system~\eqref{eq:equilibrium_internal_strain} provides the incremental periodic displacement field $\Tilde{\bq}^*$ internal to the lattice for every given $\bL$.
As a consequence of the linearity of $\hat{\bq}(\bL)$, the solution $\Tilde{\bq}^*(\bL)$ is, in turn, a linear function of $\bL$.

A few considerations have to be made about the solvability of the system~\eqref{eq:equilibrium_internal_strain}.
In fact, it is easy to show that the matrix $\trans{\bZ_0}\bK(\bP)\,\bZ_0$ is always singular, regardless of the specific lattice structure under consideration.
This is proved by considering a vector $\Tilde{\bq}^* = \bt$ defining a pure rigid-body translation and observing that $\bK(\bP)\,\bZ_0\,\bt=\bzero$, which, in turn, implies that the dimension of the nullspace of $\trans{\bZ_0}\bK(\bP)\,\bZ_0$ is \textit{at least} 2, as two linearly independent rigid-body translations exist for a 2D lattice.
Any other deformation mode, possibly contained in $\ker(\trans{\bZ_0}\bK(\bP)\,\bZ_0)$, is therefore a zero-energy mode, called `floppy mode' \cite{mao_2018,zhang_2018}.
These modes are excluded in the following analysis to ensure solvability of system~\eqref{eq:equilibrium_internal_strain}, so that $\ker(\trans{\bZ_0}\bK(\bP)\,\bZ_0)$ contains \textit{only} two
rigid-body translations.
Floppy modes can be always recovered in limits of vanishing stiffness and can eliminated or introduced playing with prestress~\cite{pellegrino_1986,pellegrino_1990}.

Having excluded floppy modes and observing that the right-hand side of Eq.~\eqref{eq:equilibrium_internal_strain} is orthogonal to $\ker(\trans{\bZ_0}\bK(\bP)\,\bZ_0)$,
\begin{equation*}
	\bt\scalp\trans{\bZ_0}\bK(\bP)\,\hat{\bq}(\bL) = 0,
\end{equation*}
for all rigid-body translations $\bt$, the solution $\Tilde{\bq}^*(\bL)$ can now be determined.

\paragraph{Match of the second-order incremental strain-energy density and determination of the incremental constitutive tensor.}
The solution of the linear system~\eqref{eq:equilibrium_internal_strain} allows the incremental displacement, Eq.~\eqref{eq:cauchy_born_hypothesis_vector}, to be expressed only in terms of the macroscopic displacement gradient $\bL$ as $\bq(\Tilde{\bq}^*(\bL),\bL)$.
Therefore, the second-order incremental strain-energy stored in a single unit cell of the lattice undergoing a macroscopic strain can be evaluated as follows
\begin{equation}
	\label{eq:strain_energy_lattice}
	\mE(\bL) = \frac{1}{2} \, \bq(\Tilde{\bq}^*(\bL),\bL) \scalp \bK(\bP)\,\bq(\Tilde{\bq}^*(\bL),\bL) \,,
\end{equation}
which is a quadratic form in $\bL$, because $\bq(\Tilde{\bq}^*(\bL),\bL)$ is linear in $\bL$.
By equating the second-order strain-energy density of the continuum
$\mW_2(\bL) = \fC[\bL] \scalp \bL/2$
to the average energy of the lattice~\eqref{eq:strain_energy_lattice}, the following equivalence condition is obtained
\begin{equation}
	\label{eq:strain_energy_density_equivalence}
	\underbrace{ \frac{1}{2}\,\fC[\bL] \scalp \bL}_{\text{Continuum}} = \underbrace{\frac{1}{\lvert\mC\lvert}\,\mE(\bL)}_{\text{Lattice}} \,,
\end{equation}
where $\lvert\mC\lvert$ is the area of the unit cell.

Finally, a calculation of the second gradient of Eq.~\eqref{eq:strain_energy_density_equivalence} with respect to $\bL$ yields the incremental constitutive tensor for the effective Cauchy material, in the form
\begin{equation}
	\label{eq:constitutive_tensor_lattice}
	\fC = \frac{1}{\lvert\mC\lvert}\,\frac{\partial^2\,\mE(\bL)}{\partial\bL\,\partial\bL}
	= \frac{1}{2\lvert\mC\lvert}\,\frac{\partial^2}{\partial\bL\,\partial\bL} \Big[ \bq(\Tilde{\bq}^*(\bL),\bL) \scalp \bK(\bP)\,\bq(\Tilde{\bq}^*(\bL),\bL) \Big] \,,
\end{equation}
which becomes now an \textit{explicit function of the prestress state}, as well as of all the mechanical parameters defining the lattice.

\paragraph{Match of the first-order incremental strain-energy density and homogenization of the prestress state.}
\label{sec:homogenization_prestress}
So far, the incremental constitutive tensor $\fC$ of a continuum `equivalent' to a prestressed elastic lattice, Eq.~\eqref{eq:constitutive_tensor_lattice}, has been obtained through homogenization.
It is important now to `dissect' from  $\fC$ the effect of the prestress $\bT$ and, as a consequence, to obtain tensor $\fE$.

It will be shown below that the current prestress state $\bT$ of the homogenized material can directly be linked to the preload state $\bP=\{P_1,...,P_{N_{\text{B}}}\}$ of the lattice.
In fact, by observing that equation~\eqref{eq:strain_energy_density_equivalence} represents the \textit{second-order} incremental strain energy, equal to $\mW_2(\bL)=\dot{\bS}(\bL)\scalp\bL/2$, an equivalence analogous to that expressed by equation~\eqref{eq:strain_energy_density_equivalence} can be obtained considering the \textit{first-order} increment of the strain energy, $\mW_1(\bL)=\bT\scalp\bL$.
Thus, the first-order term can be identified as the \textit{average work done by the prestress state} $\bP$ \textit{during the lattice deformation} $\bq(\Tilde{\bq}^*(\bL),\bL)$ \textit{induced by} $\bL$.
Accordingly, the following equivalence can be stated
\begin{equation}
	\label{eq:work_prestress_equivalence}
	\underbrace{\bT\scalp\bL}_{\text{Continuum}} = \underbrace{\frac{1}{\lvert\mC\rvert}\,\bef_\bP\scalp\bq(\Tilde{\bq}^*(\bL),\bL)}_{\text{Lattice}} \,,
\end{equation}
where vector $\bef_\bP$ collects the forces that emerge at the nodes of the unit cell and are in equilibrium with the axial preload $\bP$, \textit{in the current configuration assumed as reference}.
As a consequence, the forces $\bef_\bP$ are independent of $\bL$ and linear in $\bP$.

Equation \eqref{eq:work_prestress_equivalence} requires that the work done by axial loads $\bef_\bP$ for nodal displacements $\bq$ associated to a skew-symmetric velocity gradient $\bL = \bW$ be zero, namely
\begin{equation}
	\label{eq:work_prestress_spin}
	\bef_\bP\scalp\bq(\Tilde{\bq}^*(\bW),\bW) = 0 \,.
\end{equation}
This statement is a direct consequence of the principle of virtual work for rigid body incremental motions, because $\bq(\Tilde{\bq}^*(\bW),\bW)$ represents an incremental rotation of the lattice and $\bef_\bP$ satisfies equilibrium.
Hence, taking into account the property~\eqref{eq:work_prestress_spin}, the homogenized prestress $\bT$ can be obtained as the gradient of the equivalence condition~\eqref{eq:work_prestress_equivalence} with respect to the symmetric part of $\bL$, denoted as $\bD$,
\begin{equation}
	\label{eq:prestress_tensor_lattice}
	\bT = \frac{1}{\lvert\mC\rvert}\,\deriv{}{\bD} \Big[ \bef_\bP\scalp\bq(\Tilde{\bq}^*(\bD),\bD) \Big]\,.
\end{equation}

\section{Tensile material instabilities in a preloaded lattice}
\label{sec:grid}
%
%
The analytical framework developed in Sections~\ref{sec:prestressed_lattice}, and~\ref{sec:homogenization} is now applied to a particular lattice structure in order to showcase a concrete example of a material displaying static instabilities that are triggered by both compressive and tensile presstress states.
\begin{figure}[htbp]
	\centering
	\begin{subfigure}[b]{0.48\linewidth}
		\centering
		\caption{Unit cell}
		\includegraphics[width=0.98\linewidth]{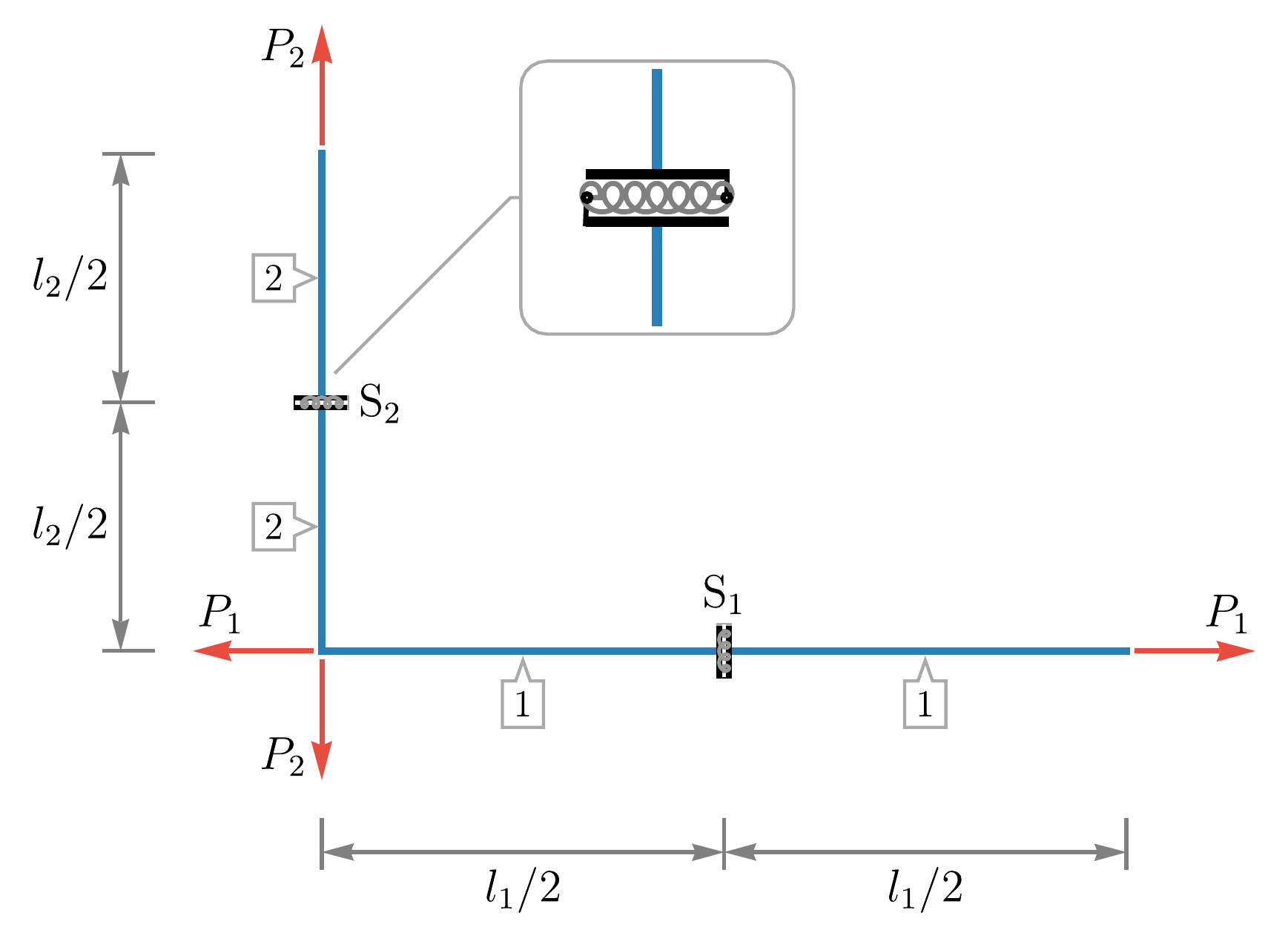}
		\label{fig:sliding_grid_unit_cell}
	\end{subfigure}
	\begin{subfigure}[b]{0.48\linewidth}
		\centering
		\caption{Periodic lattice}
		\includegraphics[width=0.98\linewidth]{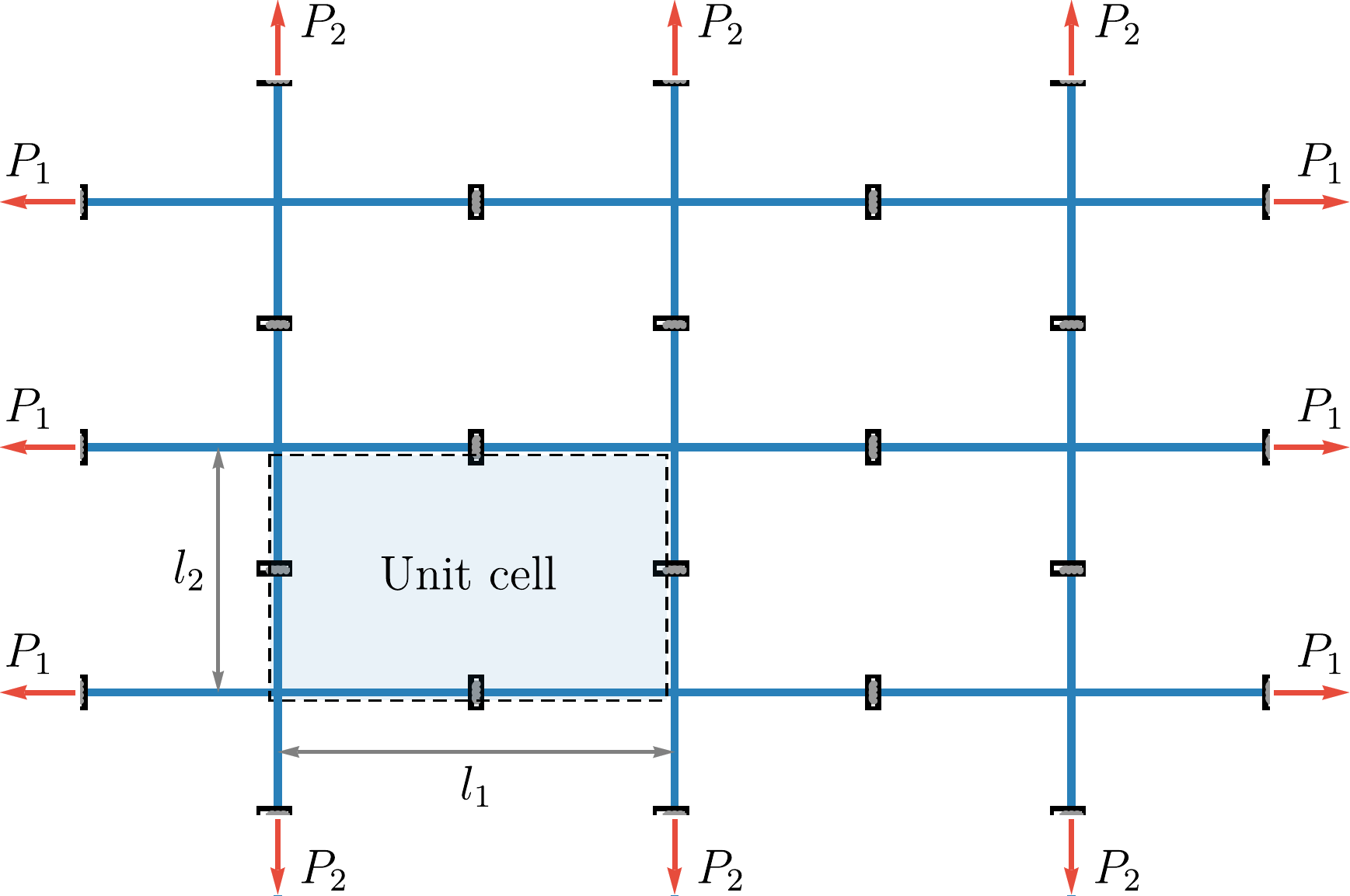}
		\label{fig:sliding_grid_lattice}
	\end{subfigure}
	\caption{
		The rectangular lattice of preloaded rods endowed with sliding constraints used to realize a material capable of losing ellipticity under tensile and compressive loadings.
		Linear springs are introduced to stiffen the sliders, thus preventing floppy modes at vanishing preload.
	}
	\label{fig:sliding_grid}
\end{figure}
The lattice under analysis is the rectangular grid illustrated in Fig.~\ref{fig:sliding_grid} which is preloaded along the two principal directions and endowed with sliding constraints.
The sliders are stiffened by linear springs to prevent trivial floppy modes that would otherwise be present when the preload vanishes.
The unit cell of the periodic structure, Fig.~\ref{fig:sliding_grid_unit_cell}, is chosen to provide the minimum number of rods and thus simplify the computations involved in the stability analysis.

The lattice configuration is parametrized by the following dimensionless ratios
\begin{equation}
	\label{eq:dimensionless_groups}
	p_i = \frac{P_i\,l_i^2}{B_i}\,, \quad \lambda_i = l_i \sqrt{\frac{A_i}{B_i}}\,, \quad \kappa_i = \frac{k_i\,l_i^3}{B_i}\,, \quad \xi = l_2/l_1\,, \quad \chi = A_2/A_1\,, \quad \forall i\in\{1, 2\}\,,
\end{equation}
where the index $i$ identifies the horizontal and vertical rods according to Fig.~\ref{fig:sliding_grid_unit_cell}, while $k_i$ denotes the sliding stiffness of the sliders provided by the linear springs.

Buckling of structures similar to those considered here and embedding sliding constraints has been considered in~\cite{eremeyev_2020}, under the hypotheses that the structure has a finite-size and is subject to equibiaxial loading.
In~\cite{eremeyev_2020} the rods are assumed as axially extensible, but rigid under bending, while both deformations are taken into account in the present work.
Moreover, an effective continuum material for such structures has not been given in~\cite{eremeyev_2020}.

The general homogenization method developed in Section~\ref{sec:homogenization}, applied to the lattice material represented in Fig.~\ref{fig:sliding_grid}, allows the identification of the effective incremental constitutive tensor $\fC$ and prestress tensor $\bT$ (which may be conveniently made dimensionless through multiplication by $l_1/A_1$) as \textit{explicit} functions of all parameters~\eqref{eq:dimensionless_groups}.
The resulting expression of the constitutive tensor $\fC$ is quite lengthy and is deferred to Appendix~\ref{sec:effective_constitutive_tensor},  while the homogenized prestress tensor $\bT$ can be compactly expressed as
\begin{equation*}
    \bT = \frac{P_1}{l_2}\,\be_1\otimes\be_1 + \frac{P_2}{l_1}\,\be_2\otimes\be_2 \,.
\end{equation*}

\subsection{Positive definiteness, strong ellipticity, and lattice stability}
\label{sec:PD_SE_LS}
%
%
A comprehensive stability analysis of the orthotropic lattice under study requires the determination of the threshold for the applied preload which triggers one or multiple non-trivial static bifurcations~\cite{bordiga_2021}.

With regards to the stability of the effective medium, the positive definiteness (PD) of the incremental constitutive operator $\fC$ ensures the uniqueness of the incremental boundary value problem subject to arbitrary boundary conditions~\cite{bigoni_2012}.
The effective material is defined PD if
\begin{equation}
	\label{eq:PD_effective_medium}
	\fC[\bL] \scalp \bL > 0 \,, \qquad \forall \bL \neq \bzero \,,
\end{equation}
which, due to the \textit{major symmetry} of $\fC$, is equivalent to the positiveness of all the eigenvalues of $\fC$.
When PD is lost at a given loading threshold, condition~\eqref{eq:PD_effective_medium} does not hold true and therefore zero-energy modes (or floppy modes) arise.
Thus if the effective material is PD, the lattice can be considered `macroscopically stable' against arbitrary disturbances.
Otherwise, some macroscopic deformation exist, which is associated to a zero energy expenditure.

Strong ellipticity (SE) characterizes the stability of the effective medium with respect to perturbations that vanish on the boundary of an arbitrary small sphere (corresponding to the so-called `van Hove problem'~\cite{bigoni_2012}) and is defined as the positive definiteness of the acoustic tensor $\bA^{(\fC)}(\bn)$ associated to the incremental fourth-order tensor $\fC$
\begin{equation}
	\label{eq:SE_effective_medium}
	\bg \scalp \bA^{(\fC)}(\bn)\,\bg > 0 \qquad \forall\bn\neq\bzero \quad \forall\bg\neq\bzero \,,
\end{equation}
where $\bA^{(\fC)}(\bn)\,\bg=\fC[\bg\otimes\bn]\,\bn$.

On the other hand, the bifurcation of an incrementally loaded periodic lattice can be analyzed with a Floquet-Bloch technique as shown in~\cite{triantafyllidis_1993,bordiga_2021}.
This analysis shows that both local and global bifurcation modes can occur and only the latter correspond to failure of ellipticity for the effective material evaluated from homogenization~\cite{pontecastaneda_1989,pontecastaneda_1997,furer_2018}.

\subsection{A bounded stability domain}
\label{sec:stability_domains}
%
%
Stability domains represent an effective tool to characterize the regions of the prestress space where a material is stable.
Their boundaries define the thresholds of instability.

Stability domains for the orthotropic lattice shown in  Fig.~\ref{fig:sliding_grid} are computed in the 2D dimensionless prestress space $\{p_1,p_2\}$ for several values of the dimensionless sliding stiffnesses $\kappa_1$ and $\kappa_2$ with the purpose of investigating both cubic and orthotropic configurations, including the limiting cases $\kappa_1\to\infty$ and $\kappa_2\to\infty$.
Physically, these limits correspond to sliders with infinite sliding stiffness, thus realizing perfect \lq welding' conditions.
Several configurations for `fully welded' grids have been explored in~\cite{bordiga_2021} where it has been shown that the stability domain of these materials is unbounded for tensile preloads.
However, the results reported in this section demonstrate that the introduction of sliding constraints strongly alters the structure of the stability domain.

For the computation of the stability domains, the slenderness values are set equal to  $\lambda_1=\lambda_2=20$, while the aspect ratio and the area ratio are chosen as $\xi=\chi=1$.

\begin{figure}[htbp]
	\centering
	{\phantomsubcaption\label{fig:stability_domains_cubic_1}}
	{\phantomsubcaption\label{fig:stability_domains_cubic_2}}
	{\phantomsubcaption\label{fig:stability_domains_cubic_5}}
	{\phantomsubcaption\label{fig:stability_domains_cubic_10}}
	{\phantomsubcaption\label{fig:stability_domains_cubic_20}}
	{\phantomsubcaption\label{fig:stability_domains_cubic_100}}
	\includegraphics[width=0.9\linewidth]{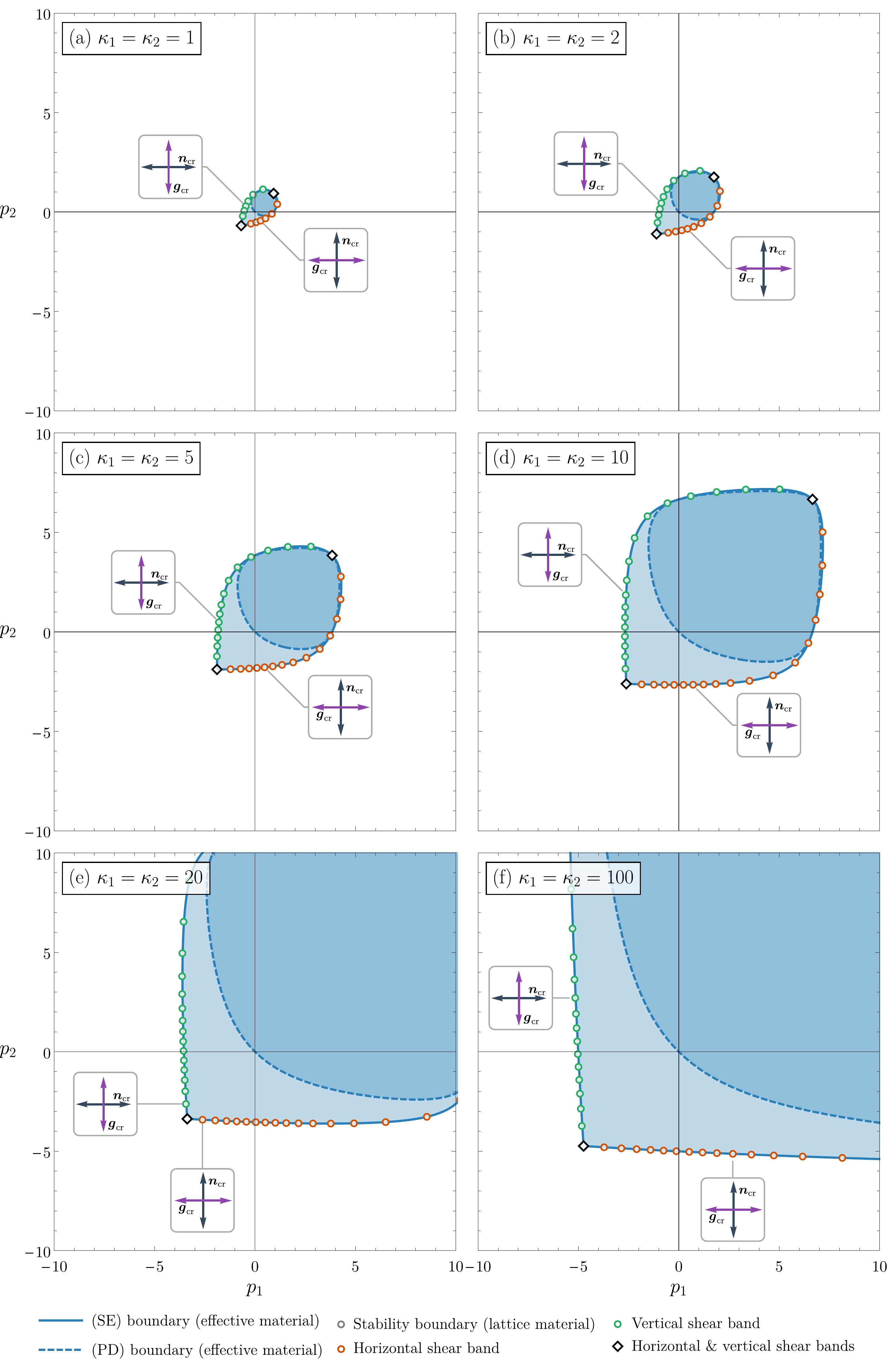}
	\caption{
		Strong ellipticity (SE), positive definiteness (PD), and lattice stability domains for the \textit{cubic} grid with $\lambda_1=\lambda_2=20$, $\xi=\chi=1$, and six values of sliding stiffness $\kappa_1=\kappa_2=1,2,5,10,20,100$.
		The arrows sketched in the insets represent the critical direction $\bn_{\text{cr}}$ and the associated mode $\bg_{\text{cr}}$ responsible for the loss of strong ellipticity.
	}
	\label{fig:stability_domains_cubic}
\end{figure}

\begin{figure}[htbp]
	\centering
	{\phantomsubcaption\label{fig:stability_domains_orthotropic_1_1}}
	{\phantomsubcaption\label{fig:stability_domains_orthotropic_1_2}}
	{\phantomsubcaption\label{fig:stability_domains_orthotropic_1_5}}
	{\phantomsubcaption\label{fig:stability_domains_orthotropic_1_10}}
	{\phantomsubcaption\label{fig:stability_domains_orthotropic_1_20}}
	{\phantomsubcaption\label{fig:stability_domains_orthotropic_1_100}}
	\includegraphics[width=0.9\linewidth]{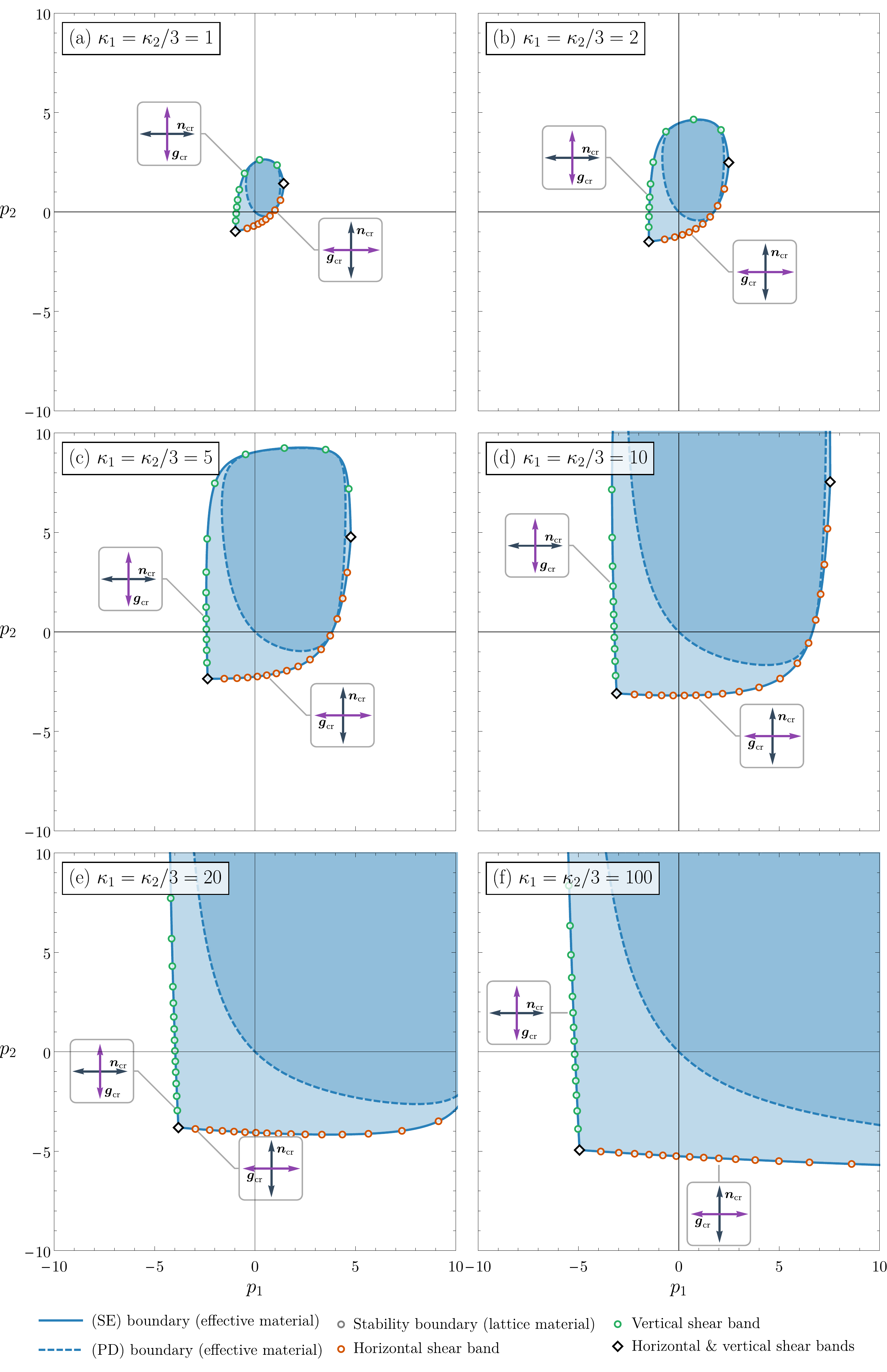}
	\caption{
		Strong ellipticity (SE), positive definiteness (PD), and lattice stability domains for the \textit{orthotropic} grid with $\lambda_1=\lambda_2=20$, $\xi=\chi=1$, and six values of sliding stiffness $\kappa_1=\kappa_2/3=1,2,5,10,20,100$.
		Compared to the cubic case reported in  Fig.~\ref{fig:stability_domains_cubic}, the orthotropy induced by the different sliding stiffnesses increases  the size of the stability domain along the direction of the stiffest sliders (vertical).
	}
	\label{fig:stability_domains_orthotropic_1}
\end{figure}

\begin{figure}[htbp]
	\centering
	\includegraphics[width=0.9\linewidth]{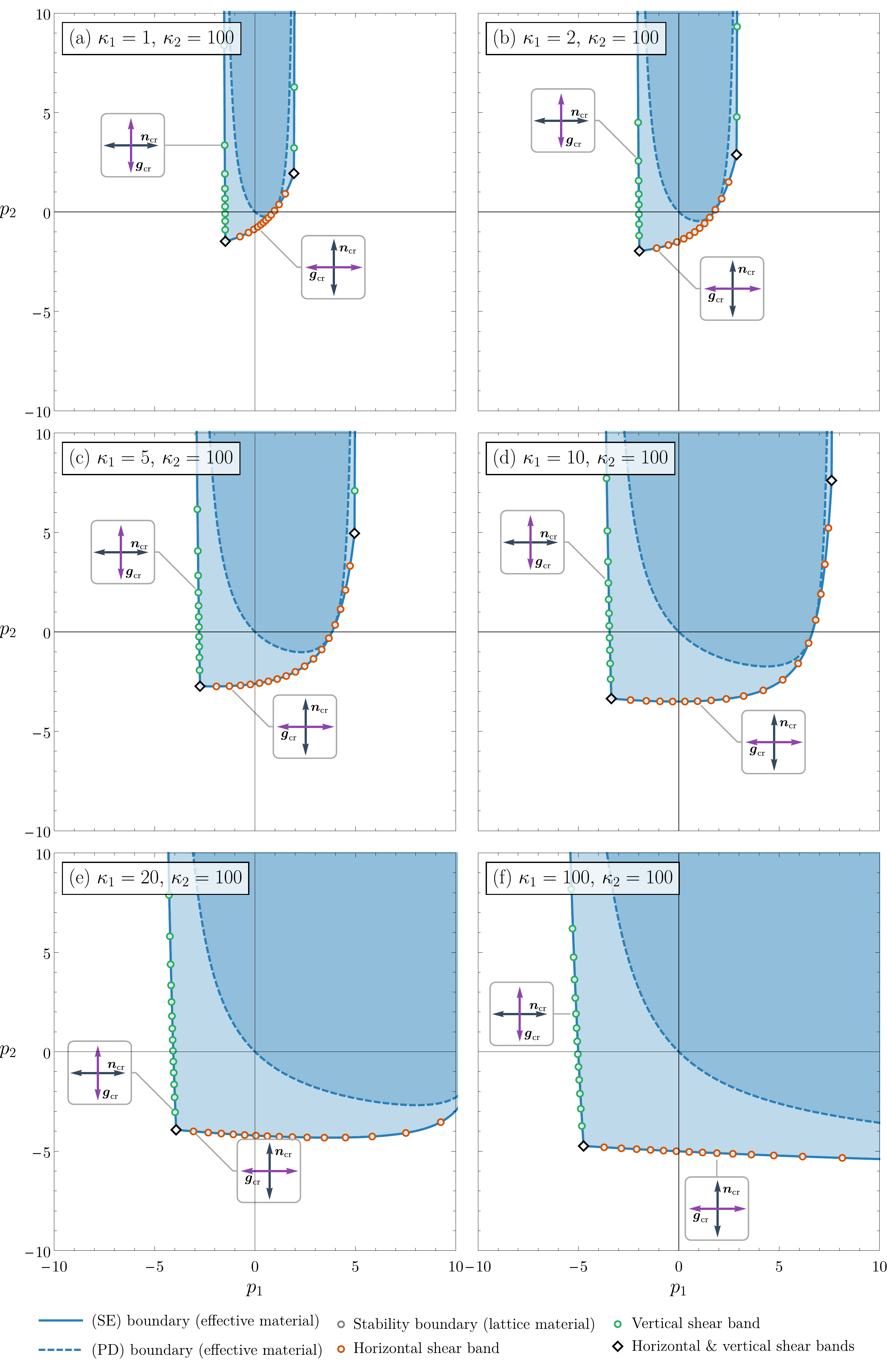}
	\caption{
		As for Fig.~\ref{fig:stability_domains_orthotropic_1}, but for values of sliding stiffness $\kappa_1=1,2,5,10,20,100$ and $\kappa_2=100$, representing the case of a strongly orthotropic grid with an almost unbounded stability domain for vertical tensile loading.
	}
	\label{fig:stability_domains_orthotropic_2}
\end{figure}

Results are reported in Figs.~\ref{fig:stability_domains_cubic}--\ref{fig:stability_domains_orthotropic_2}.
Each figure contains six plots corresponding to different values of sliding stiffness, and in each plot three kinds of stability domains are illustrated.
The shaded regions in dark blue and bounded by a dashed line represent preload states where the effective material is PD, while the light blue regions bounded by a solid line define the domain where the effective material is SE.
Note that as a consequence of Eqs.~\eqref{eq:PD_effective_medium} and~\eqref{eq:SE_effective_medium} PD regions are always \textit{contained} inside SE regions.
The third domain is the one enclosed by the colored spots, which define the region where the lattice is stable, so that both long and short-wavelength bifurcations are excluded.

It can be observed in all of the three Figs.~\ref{fig:stability_domains_cubic}--\ref{fig:stability_domains_orthotropic_2} that the boundary for lattice stability (colored spots) coincides with the boundary of SE for the effective material and encloses the origin.
This implies that the critical bifurcation occurring in the lattice is \textit{macroscopic}, namely, a long-wavelength bifurcation corresponding to the formation of a \textit{shear band}.

The red and green spots denote the presence of a horizontal and vertical shear band, respectively, while the insets depict the corresponding critical dyad  $\bn_{\text{cr}}\otimes\bg_{\text{cr}}$ responsible for the loss of SE along the direction $\bn_{\text{cr}}$ and with critical polarization $\bg_{\text{cr}}$.
Note also that the diamond-shaped spots (two for each domain) denote limit points characterized by the \textit{simultaneous} occurrence of a vertical and a horizontal shear band.

Fig.~\ref{fig:stability_domains_cubic} shows stability domains for different slider stiffness $\kappa_1=\kappa_2 = \{1, 2, 5, 10, 100\}$ of a lattice with cubic symmetry.
Figs.~\ref{fig:stability_domains_orthotropic_1} and \ref{fig:stability_domains_orthotropic_2} refer to $\kappa_1 \neq \kappa_2$ and therefore refer to orthotropy.
In the case of cubic symmetry, Fig.~\ref{fig:stability_domains_cubic}, the equibiaxial loading path $p_1=p_2$ becomes a symmetry axis for the stability regions, while this symmetry is broken when the two sliders have a different stiffness, $\kappa_1\neq\kappa_2$,  Figs.~\ref{fig:stability_domains_orthotropic_1},~\ref{fig:stability_domains_orthotropic_2}.

\textit{Remarkably, the stability domain of the lattice is bounded both in compression and in tension}, and tends to become unbounded in tension when the stiffness of the sliders increases, so that in the limit
$\kappa_1\to\infty$ and $\kappa_2\to\infty$ the case of a \lq welded connection' is recovered for which there is no bifurcation in tension.
In addition, three non-trivial features of the stability domains can be highlighted:
\begin{enumerate}[label=(\roman*)]
	\item The SE and PD limit points \textit{coincide for uniaxial tension} loading along the horizontal ($p_2=0$) and vertical direction ($p_1=0$).
	      This is clearly visible in Figs.~\ref{fig:stability_domains_cubic_1}--\subref{fig:stability_domains_cubic_10} and Figs.~\ref{fig:stability_domains_orthotropic_1_1}--\subref{fig:stability_domains_orthotropic_1_10}.
	\item The transition from a single horizontal (occurring for $p_2 < p_1$) to a single vertical (occurring for $p_2 > p_1$) shear band is marked by the equibiaxial loading path $p_1=p_2$  \textit{regardless of the values of sliding stiffness} $\kappa_1$ and $\kappa_2$. 
	      This can be observed in  Figs.~\ref{fig:stability_domains_cubic}--\ref{fig:stability_domains_orthotropic_2} by noting that the two limit points, leading to the formation of two shear bands and marked with diamond-shaped spots in the figures, lie on the equibiaxial line.
	      Note also that these points are corner points of the otherwise smooth stability boundary.
	\item In Figs.~\ref{fig:stability_domains_cubic}--\ref{fig:stability_domains_orthotropic_2}, the region corresponding to uniqueness with respect to global bifurcations in the grid of rods has a `leaf-shaped' shape and is bounded by a closed line marked with green and red spots, denoting occurrence of macroscopic bifurcations.
	      Outside this region, local bifurcations occur in the lattice, so that uniqueness is possible only \textit{inside} the `leaf-shaped' boundary.

	      In the same figures, the regions corresponding to uniqueness for the grid of rods mark the first SE boundary that is encountered in a radial (increasing) stress path by the effective material.
	      Thus, it is clear that loss of uniqueness at the `leaf-shaped' boundary coincides with the critical loss of SE in the effective material, which correctly captures instability of the elastic grid.

	      However, differently from the lattice, the effective material evidences zones outside the boundary of global instability, where SE and even PD are recovered (not shown in Figs.~\ref{fig:stability_domains_cubic}--\ref{fig:stability_domains_orthotropic_2}, but investigated in Section~\ref{sec:restab}).
	      In these zones the response of the homogenized material returns to be stable, but this stability does not reflect the behavior of the grid of rods, which is subject to local instabilities, so that the homogenization scheme does not work properly.
	      This sort of `re-stabilization' for the effective solid, which recovers SE and even PD after having lost both in a radial path of increasing prestress, is analyzed in Section \ref{sec:restab}.
\end{enumerate}

\subsection{Zero-energy modes at loss of PD of the effective material}
%
%
The stability analysis presented in Section~\ref{sec:stability_domains} demonstrates that the bounded stability domain of the lattice material endowed with sliders is correctly captured by the SE domain of the effective medium, which allows the prediction of shear band formation.

\begin{figure}[htbp]
	\centering
	\begin{subfigure}{0.8\linewidth}
		\centering
		\caption{}
		\includegraphics[width=\linewidth]{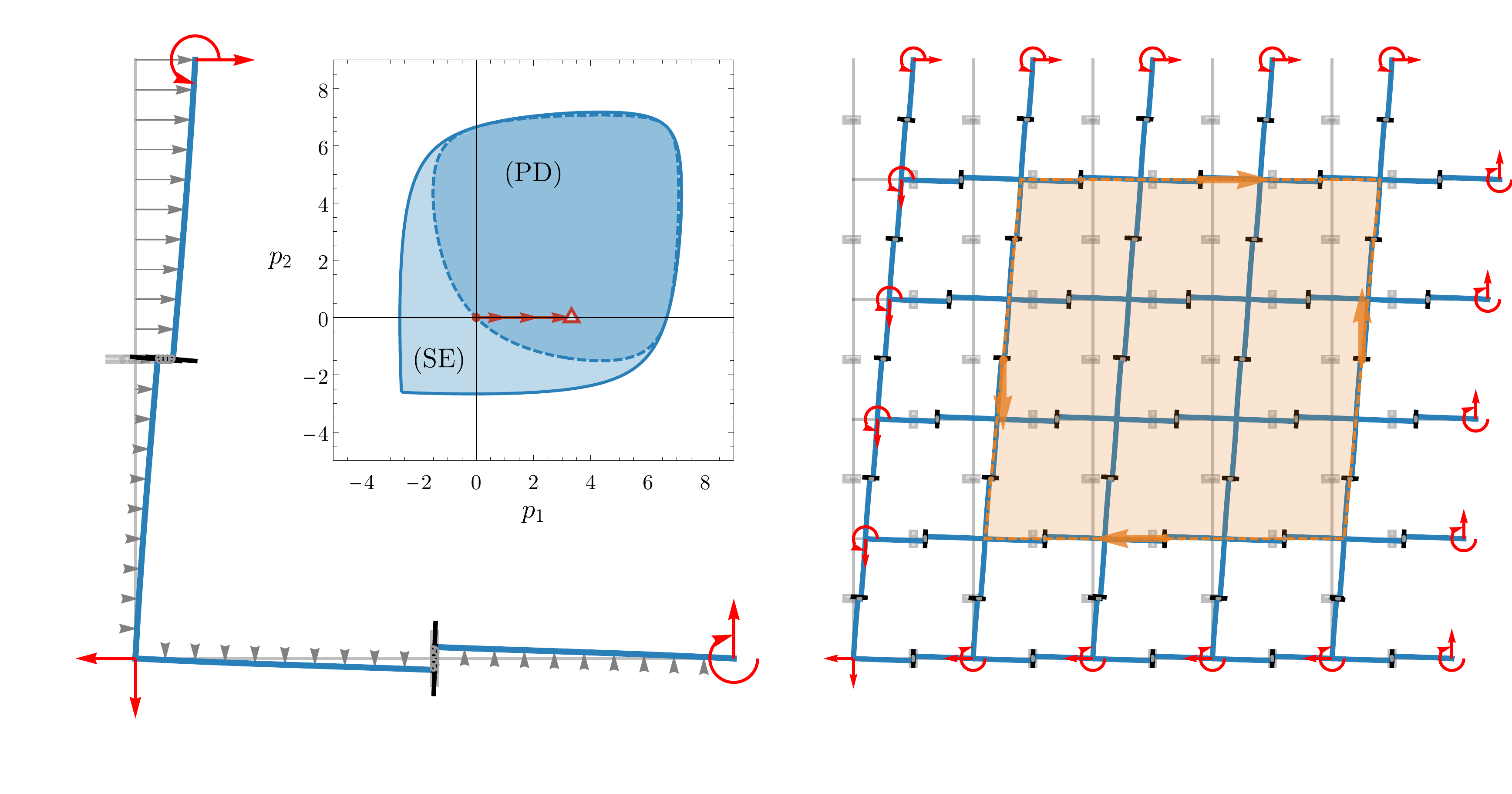}
		\label{fig:critical_mode_PD_0deg_1}
	\end{subfigure}
	\begin{subfigure}{0.8\linewidth}
		\centering
		\caption{}
		\includegraphics[width=\linewidth]{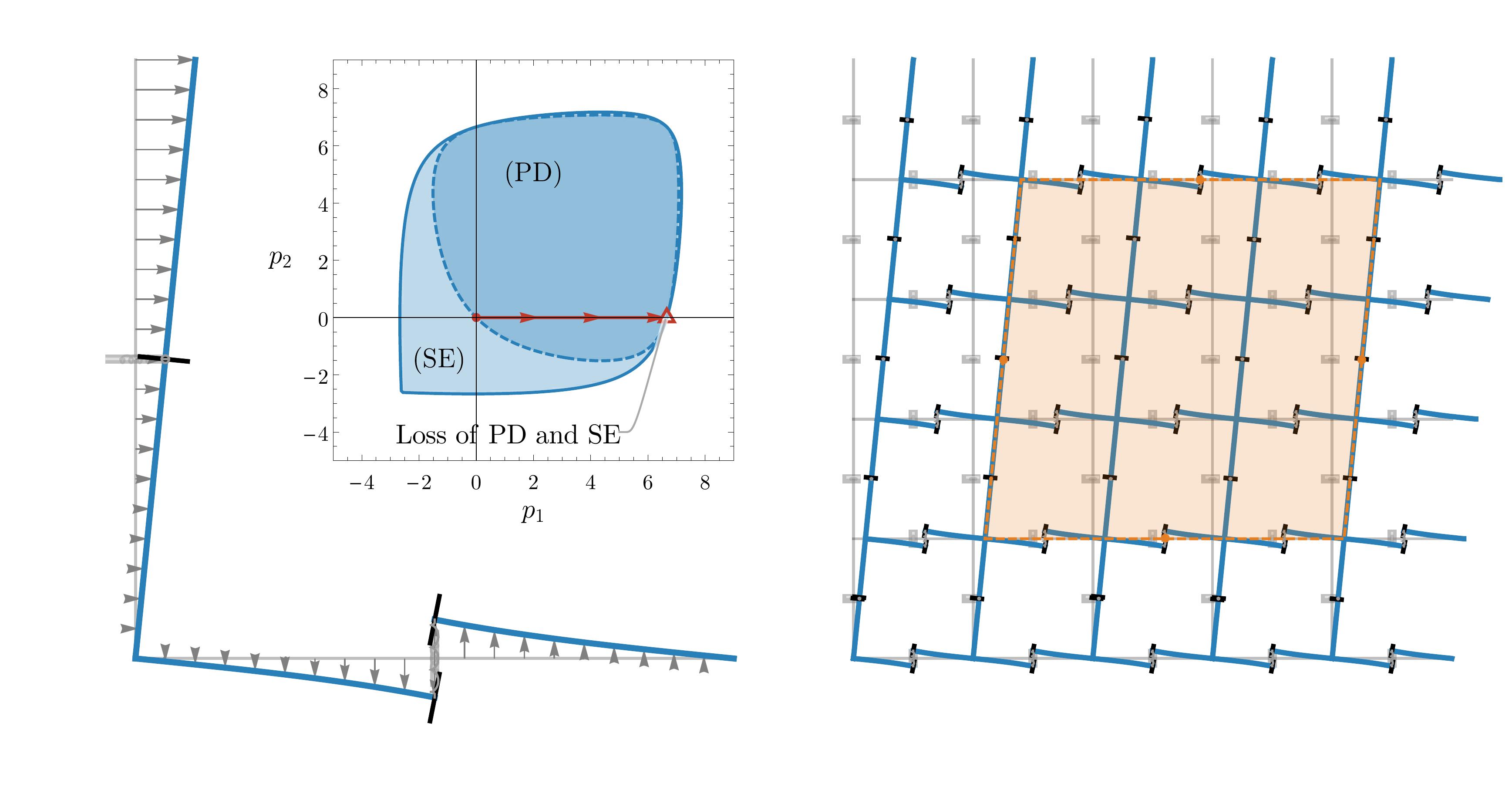}
		\label{fig:critical_mode_PD_0deg_2}
	\end{subfigure}
	\caption{
		An incremental simple shear deformation, $\bL = \be_1\otimes\be_2$, is superimposed
		upon a uniaxial prestress state.
		The response of the unit cell, together with the prestress position in the stability domain, is shown on the left, while the incremental deformation of the grid, with superimposed the incremental deformation of the effective continuum, is shown on the right.
		(\subref{fig:critical_mode_PD_0deg_1}) The prestress state is far from the PD and SE boundaries, $\bp\approx\{3.32955,0\}$.
		(\subref{fig:critical_mode_PD_0deg_2}) The prestress state belongs to both the PD and SE boundaries (coinciding for uniaxial stress), $\bp_{\text{cr}}\approx\{6.65910,0\}$; in this case, the reported incremental deformation is a zero-energy mode for the effective material.
	}
	\label{fig:critical_modes_PD}
\end{figure}

\begin{figure}[htbp]
	\centering
	\begin{subfigure}{0.8\linewidth}
		\centering
		\caption{}
		\includegraphics[width=\linewidth]{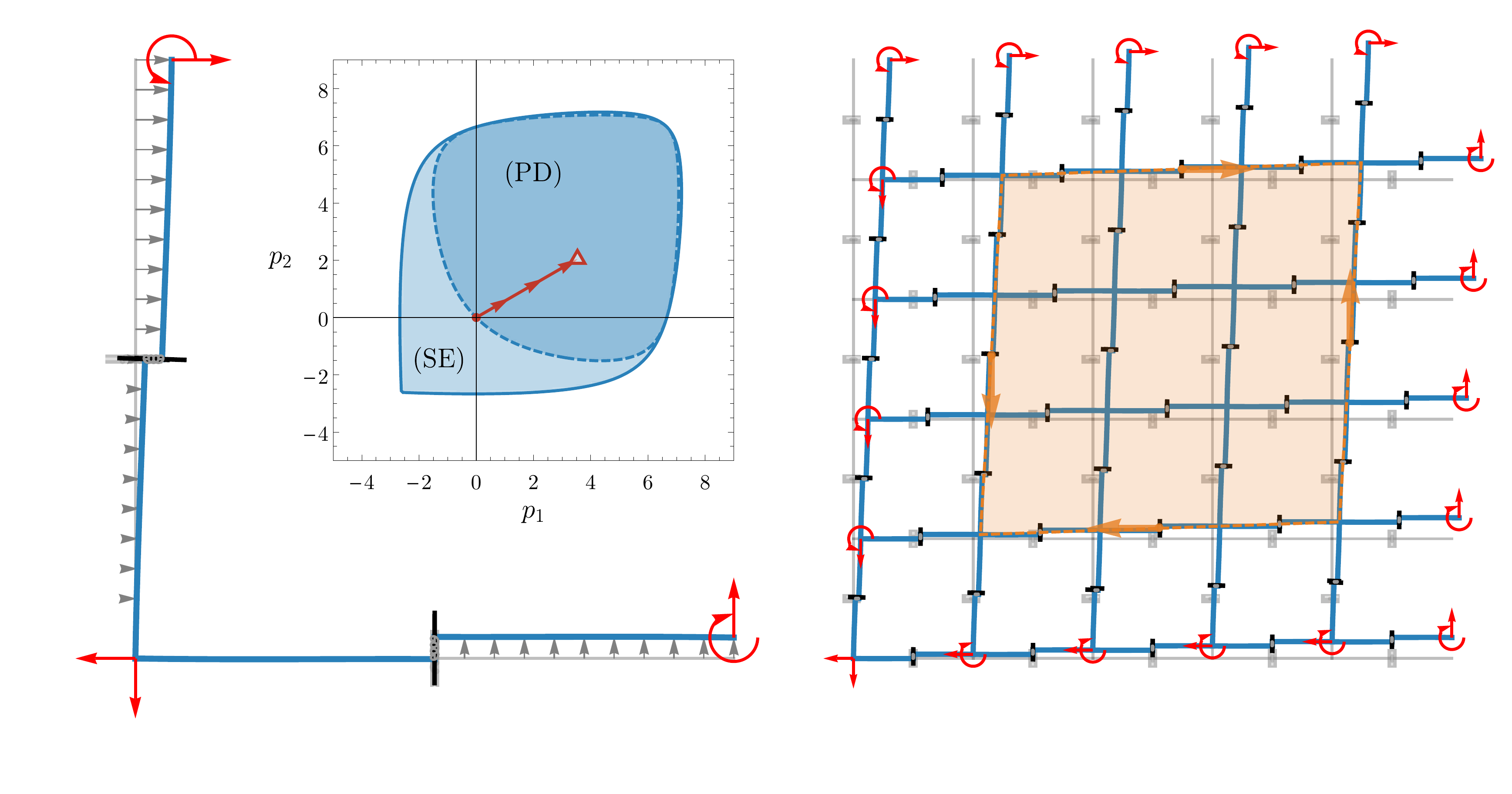}
		\label{fig:critical_mode_PD_30deg_1}
	\end{subfigure}
	\begin{subfigure}{0.8\linewidth}
		\centering
		\caption{}
		\includegraphics[width=\linewidth]{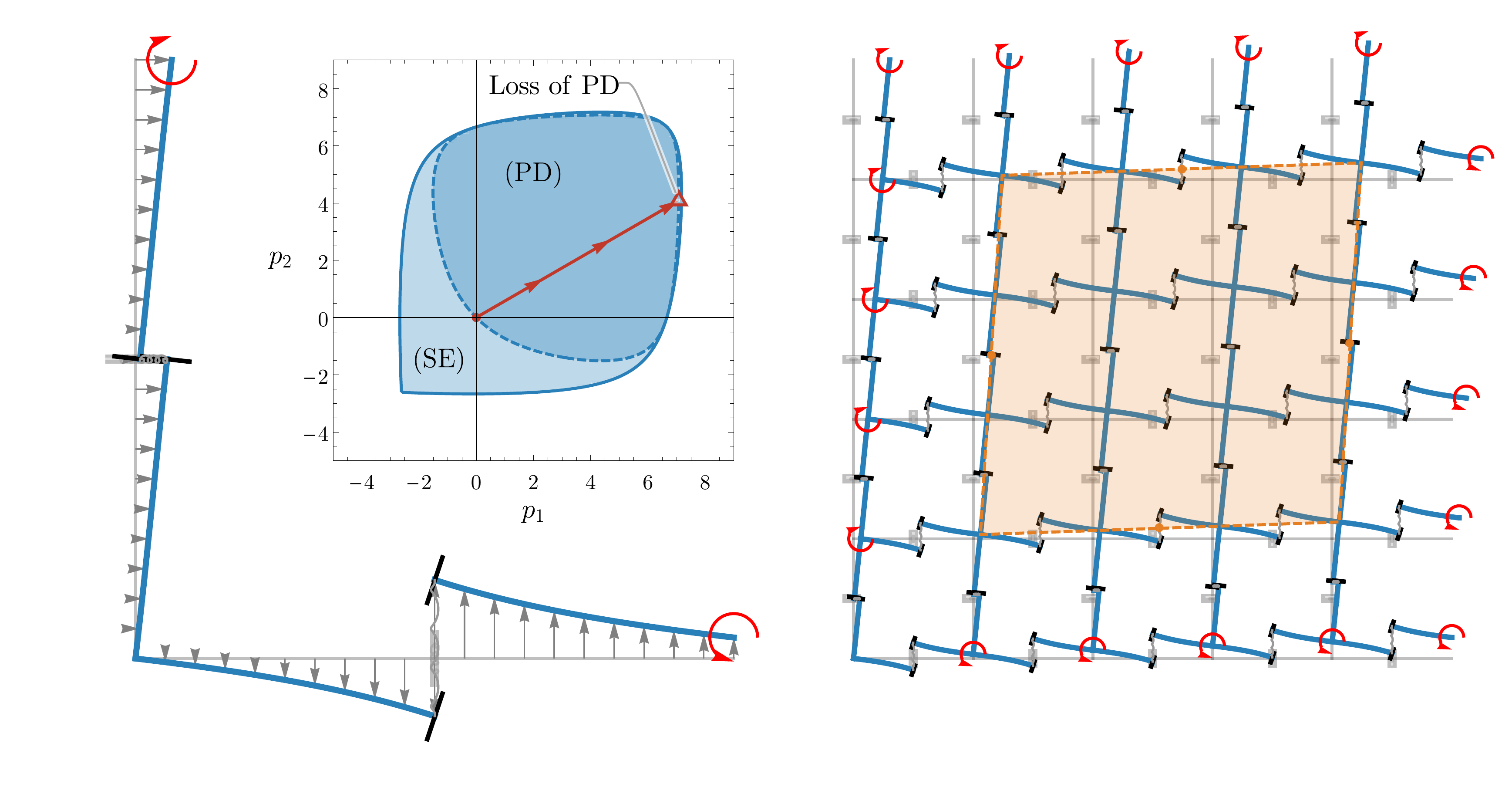}
		\label{fig:critical_mode_PD_30deg_2}
	\end{subfigure}
	\caption{As for Fig.~\ref{fig:critical_modes_PD}, except that the prestress states are tensile and biaxial, located on a line inclined at $30^\circ$ with respect to $p_1$-axis and that the superimposed deformation is an incremental deformation $\bL=0.866025\,\be_1\otimes\be_2+0.5\,\be_2\otimes\be_1$.
		(\subref{fig:critical_mode_PD_30deg_1}) The prestress is well inside the SE domain, $\bp\approx\{3.53705,2.04212\}$;
		(\subref{fig:critical_mode_PD_30deg_2}) The prestress state is on the PD boundary, $\bp_{\text{cr}}\approx\{7.07411,4.08424\}$, now inside the SE boundary, and the reported incremental deformation represents the corresponding zero-energy mode for the effective material.
	}
	\label{fig:critical_modes_PD2}
\end{figure}

An investigation on the deformation of the lattice and the effective material is presented in Figs.~\ref{fig:critical_modes_PD} and \ref{fig:critical_modes_PD2}, both referring to the following parameters of the grid
\begin{equation*}
	\lambda_1=\lambda_2=20 \,, \qquad
	\xi=\chi=1 \,, \qquad
	\kappa_1=\kappa_2=10 \,,
\end{equation*}
corresponding to one of the cases analyzed in Fig.~\ref{fig:stability_domains_cubic}.

In both figures, the grid and the effective material are prestressed up to a point denoted with a red triangle in the prestress space (inset of the figure) and then subject to an incremental deformation defined by the tensor $\bL$.
The applied incremental deformation $\bL$ is chosen to be the macroscopic zero-energy mode corresponding to the given radial loading path.
In turn, the macroscopic zero-energy modes are obtained by evaluating, on the PD boundary, the eigenvectors $\bL_{\text{cr}}$ leading to failure of the PD condition~\eqref{eq:PD_effective_medium}, or second-order work $\fC[\bL_{\text{cr}}]\scalp\bL_{\text{cr}} = 0$.
Then, by solving Eq.~\eqref{eq:equilibrium_internal_strain} and using Eq.~\eqref{eq:cauchy_born_hypothesis_vector}, the actual displacement field of the lattice can be determined and visualized.
Fig.~\ref{fig:critical_modes_PD} refers to a uniaxial tensile prestress state upon which an incremental simple shear $\bL= \be_1 \otimes \be_2$ is applied.
Fig.~\ref{fig:critical_modes_PD2} refers to a biaxial stress path inclined at $30^\circ$ with respect to the horizontal axis and an incremental deformation $\bL = 0.866025\, \be_1 \otimes \be_2 + 0.5\, \be_2 \otimes \be_1$.

It is worth noting that the zero-energy mode for uniaxial tension (Fig.~\ref{fig:critical_mode_PD_0deg_2}) highlights the peculiar interplay, occurring at the instability threshold, between the local deformation of the unit cell and the macroscopic deformation.
In fact the bifurcation mode shows the sliders within each cell opening \textit{vertically} while enabling an overall \textit{horizontal} macroscopic shearing.
Note also that under uniaxial tension, loss of positive definiteness PD for the homogenized incremental constitutive operator coincides with loss of strong ellipticity SE.

Similarly to Fig.~\ref{fig:lattice_configurations}, Figs.~\ref{fig:critical_modes_PD} and \ref{fig:critical_modes_PD2} report both the incrementally deformed grid and the corresponding incrementally deformed effective material (shown in orange).
From both figures the following conclusions can be drawn:

\begin{itemize}
	\item The bending moments applied at the ends of the rods correspond to a null mean stress and therefore do not provide any effect on the boundary of the effective material.
	      In fact, this effect is a higher-order contribution within the homogenization scheme adopted here, so that it could be highlighted only in a higher-order description employing Cosserat or Mindlin continua for the effective material.
	\item At loss of PD the effective material admits incremental deformations corresponding to zero second-order energy.
	      For these deformations (addressed in the figures), an element of the effective material is subject to an incremental strain with null surface tractions.
	      In this case, the grid of rods deforms only under incremental bending moments, which, although contributing to the local equilibrium of the lattice, do not appear on the effective continuum.
\end{itemize}
%

\section{Re-stabilization of the effective continuum induced by lattice periodic microinstabilities}
\label{sec:restab}
\begin{figure}[htbp]
	\centering
	\begin{subfigure}[t]{0.48\linewidth}
		\centering
		\caption{Stability domains beyond the first bifurcation}
		\includegraphics[width=\linewidth]{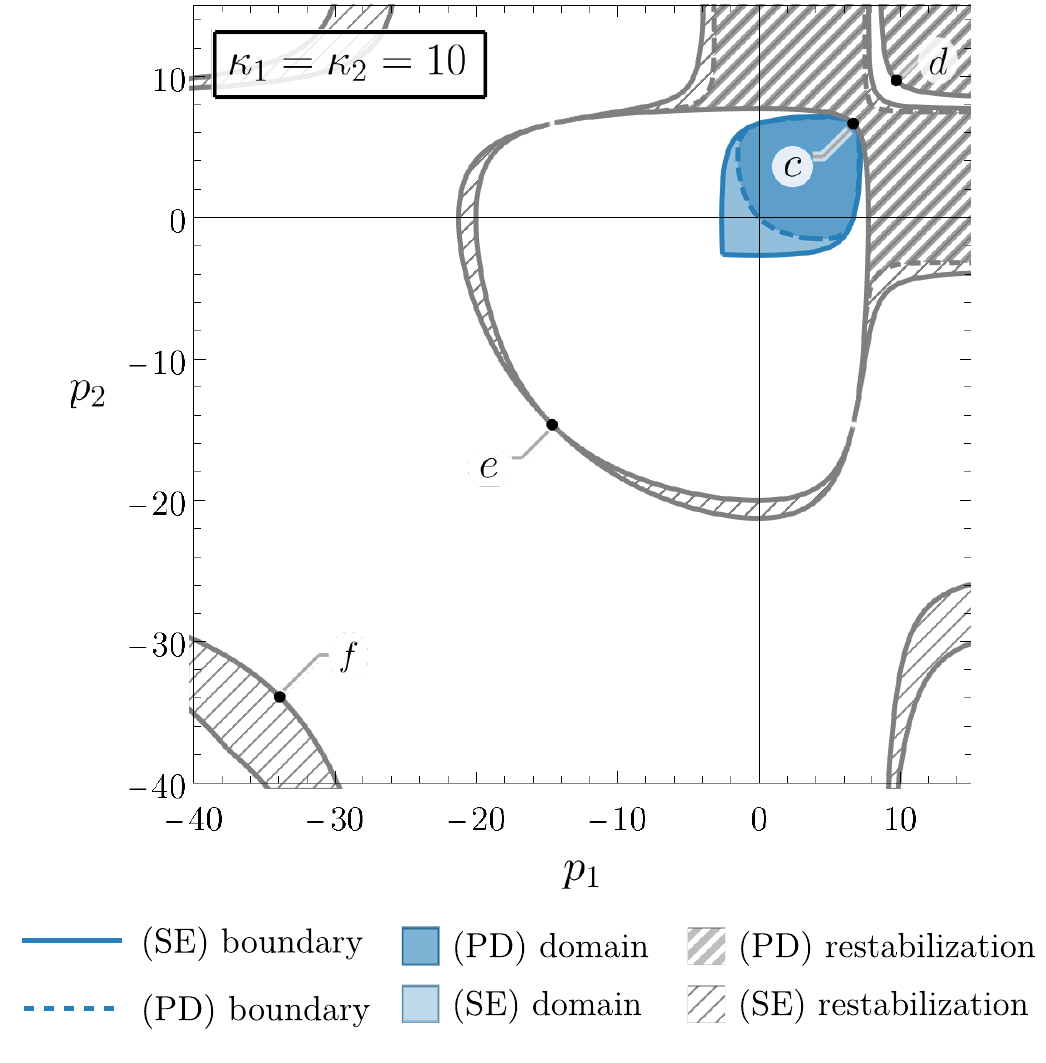}
		\label{fig:domain_restabilization}
	\end{subfigure}
	\begin{subfigure}[t]{0.48\linewidth}
		\centering
		\caption{Buckling modes of a single elastic rod}
		\includegraphics[width=\linewidth]{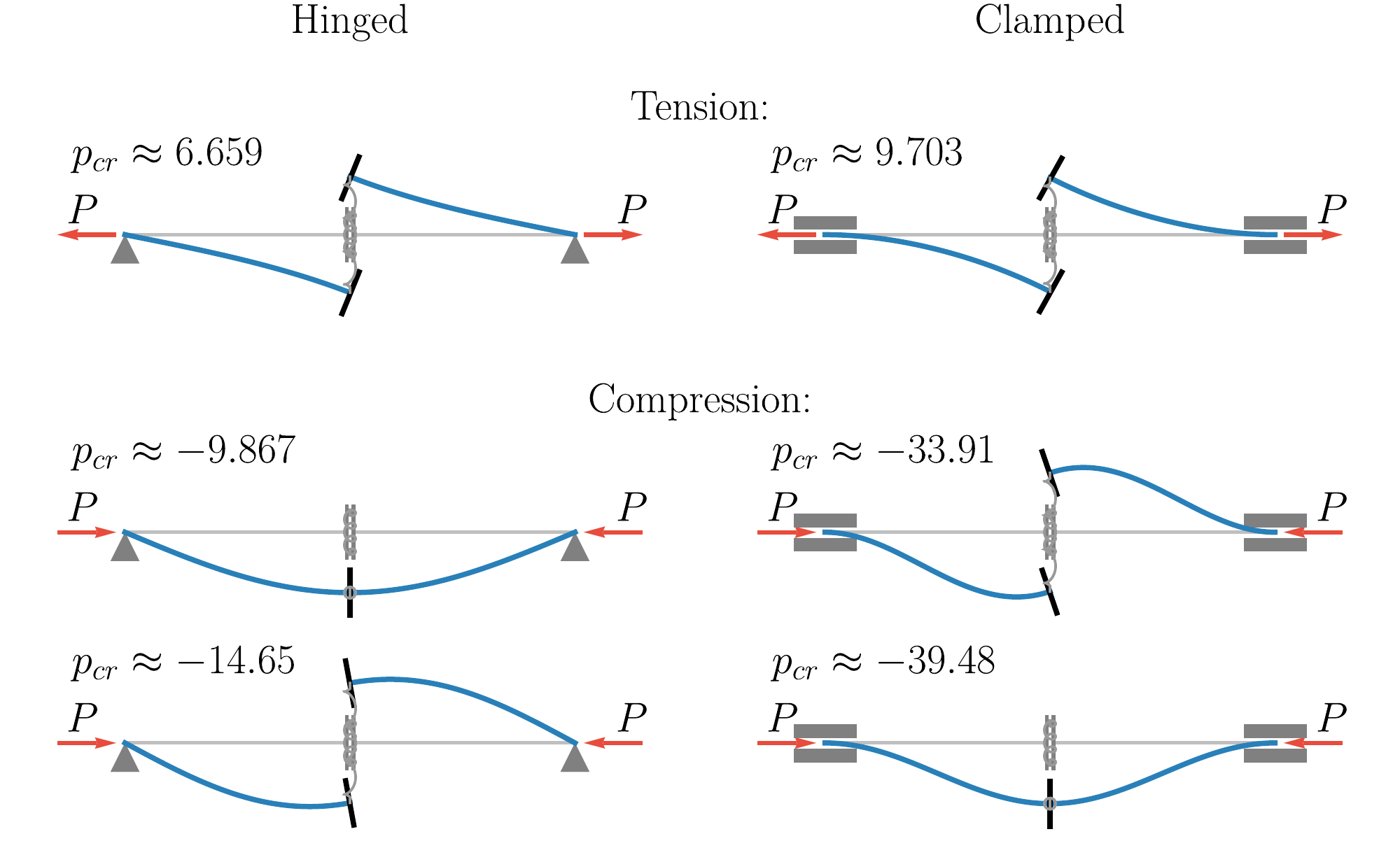}
		\label{fig:buckling_modes}
	\end{subfigure}
	\begin{subfigure}{0.48\linewidth}
		\centering
		\caption{Equibiaxial tension: $p_1=p_2 \approx 6.659$}
		\includegraphics[width=\linewidth]{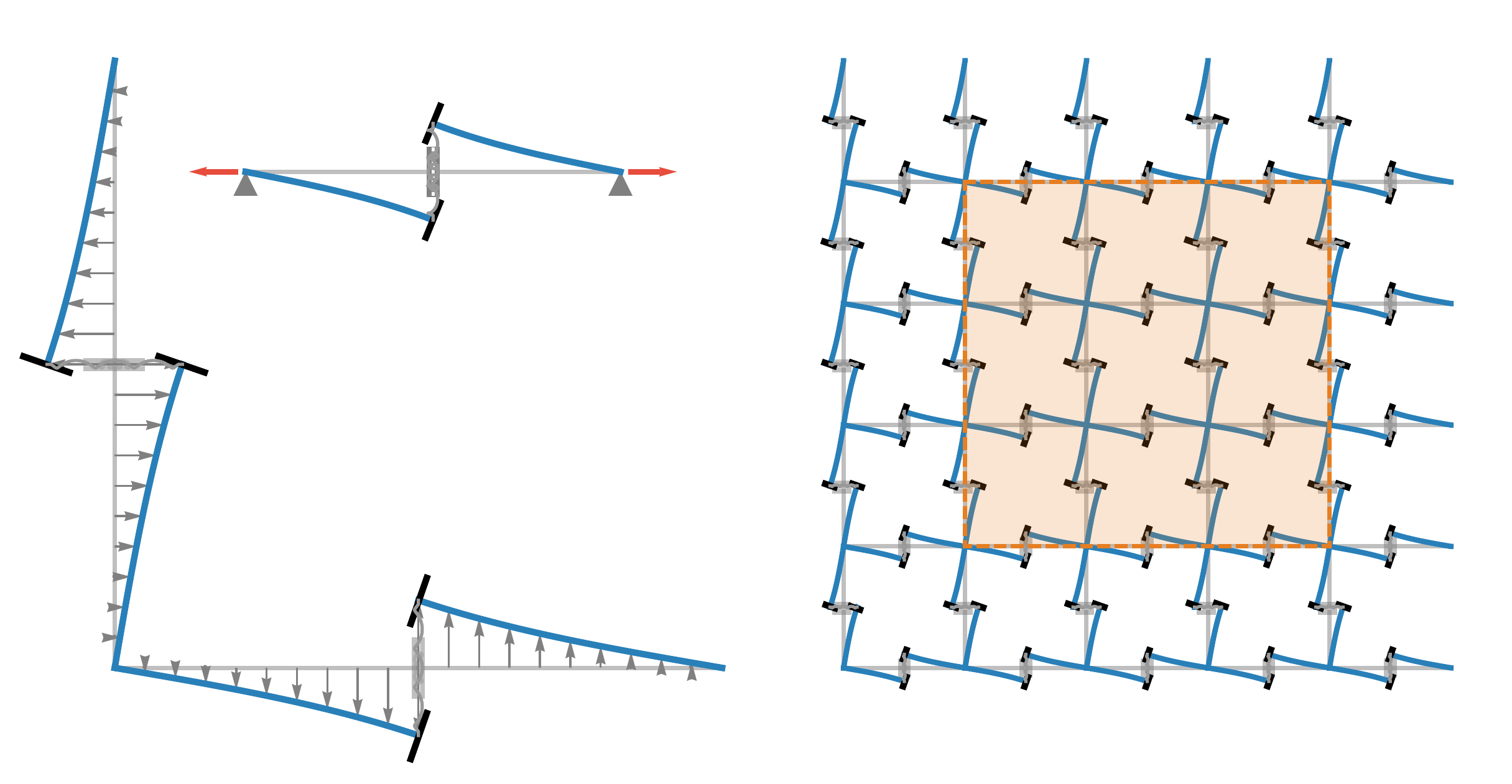}
		\label{fig:restabilization_mode_1}
	\end{subfigure}
	\begin{subfigure}{0.48\linewidth}
		\centering
		\caption{Equibiaxial tension: $p_1=p_2 \approx 9.703$}
		\includegraphics[width=\linewidth]{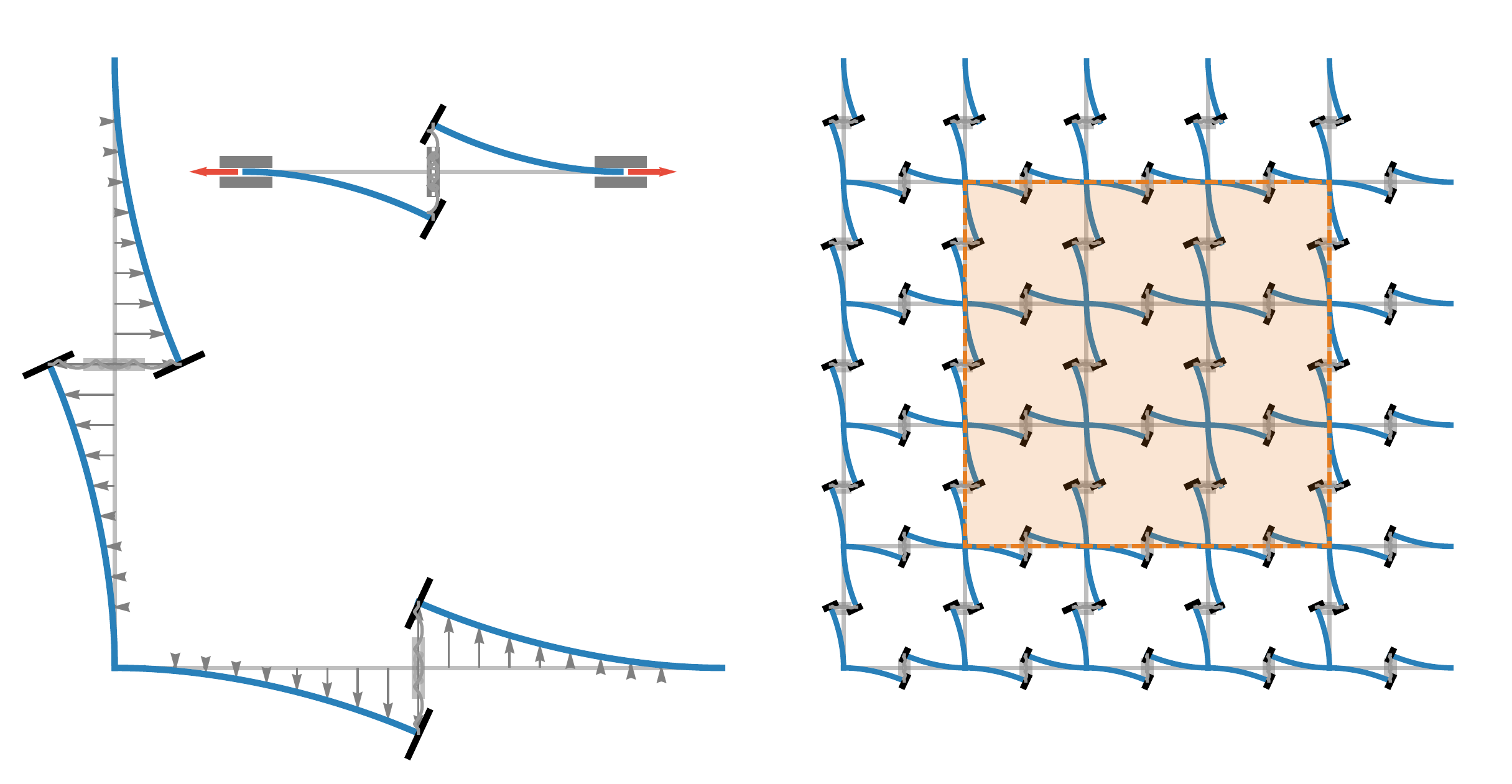}
		\label{fig:restabilization_mode_2}
	\end{subfigure}
	\begin{subfigure}{0.48\linewidth}
		\centering
		\caption{Equibiaxial compression: $p_1=p_2 \approx -14.65$}
		\includegraphics[width=\linewidth]{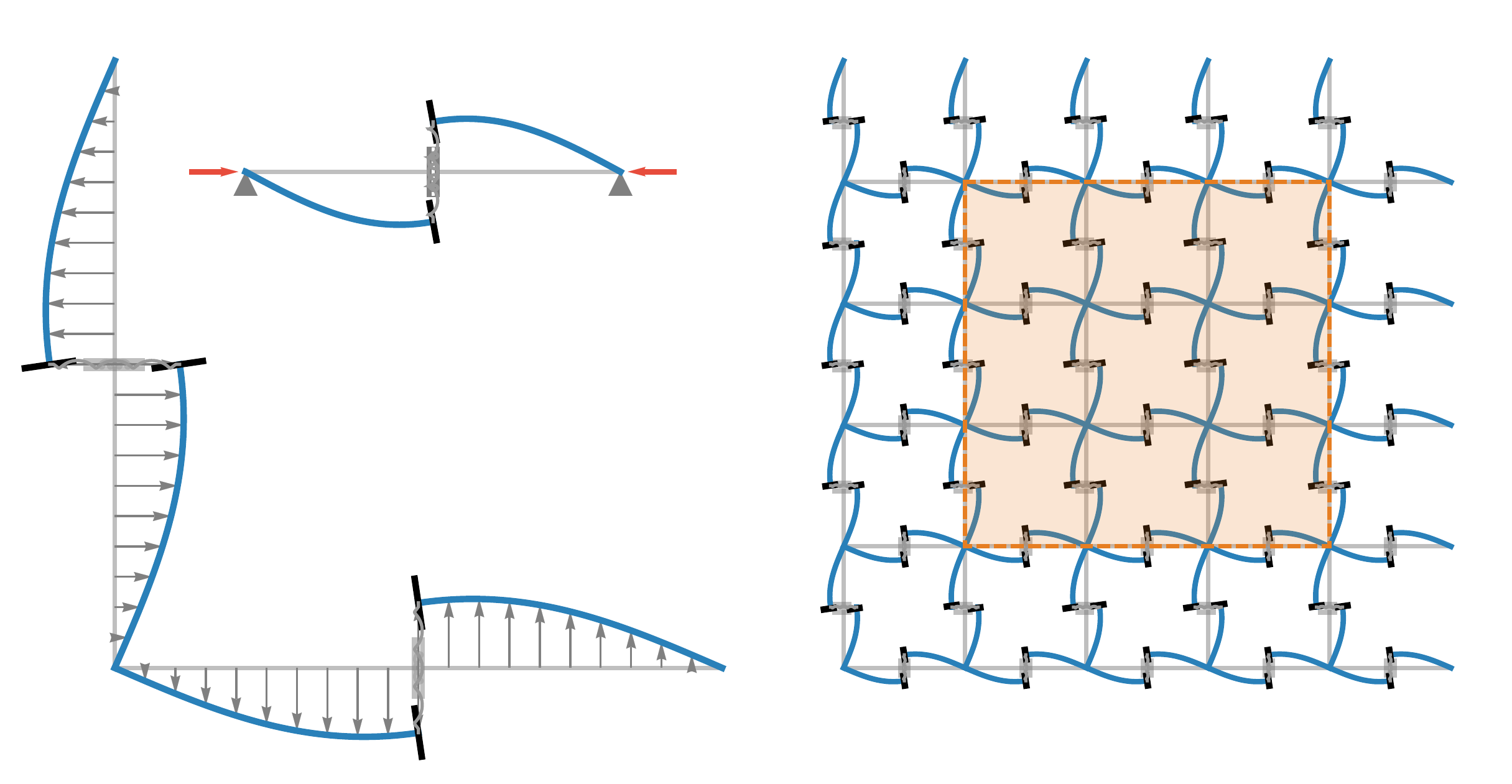}
		\label{fig:restabilization_mode_3}
	\end{subfigure}
	\begin{subfigure}{0.48\linewidth}
		\centering
		\caption{Equibiaxial compression: $p_1=p_2 \approx -33.91$}
		\includegraphics[width=\linewidth]{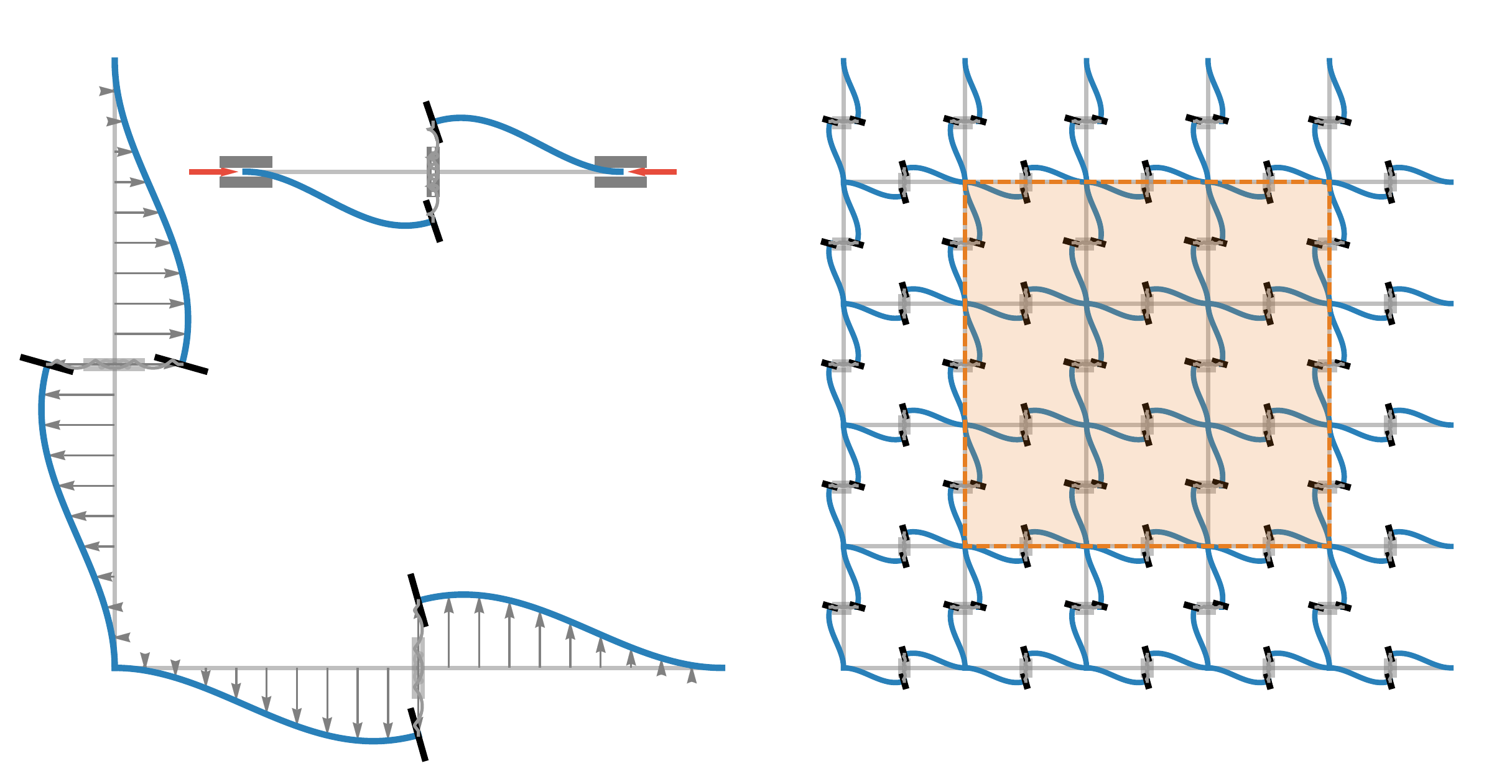}
		\label{fig:restabilization_mode_7}
	\end{subfigure}
	\caption{
		Re-stabilization of the effective material outside the domain of initial SE.
		The latter is reached when the material is subject to a radial increase in stress, but beyond this threshold, the material recovers SE and even PD.
		In these re-stabilization regions, homogenization does not work, as the grid of elastic rods is always subject to local bifurcations and is therefore unstable.  (a) Stability domains beyond the first bifurcation. (b) Buckling modes of a single elastic rod endowed with a slider and a stiffening spring: the tensile and the first two compressive bifurcation modes are shown for both hinged and clamped configurations. Lattice microbifurcations inducing re-stabilization of the effective material: (c) equibiaxial tension: $p_1 = p_2 \approx 6.659$, (d) equibiaxial tension: $p_1 = p_2 \approx 9.703$, (e) equibiaxial compression: $p_1 = p_2 \approx -14.65$ and (f) equibiaxial compression: $p_1 = p_2 \approx -33.91$.
	}
	\label{fig:restabilization}
\end{figure}

The occurrence of the `re-stabilization' mentioned in Section \ref{sec:stability_domains} is highlighted in Fig.~\ref{fig:restabilization}.
Here zones of PD (shaded dark gray) and SE (shaded light gray) are shown that extend beyond the leaf-shaped zone corresponding to the critical (in other words, the first encountered in a radial stress path) loss of SE (shaded blue) for the effective material.

This occurrence can be explained on the basis of the homogenization scheme described in Section \ref{sec:homogenization}.
One of the key points of this procedure is the solution of Eq.~(\ref{eq:equilibrium_internal_strain}), so that to obtain the periodic displacement vector $\Tilde{\bq}^*$ caused by a uniform deformation gradient $\bL$ applied to the nodes of the grid.
The solvabilty of Eq.~(\ref{eq:equilibrium_internal_strain}) relies on the absence of so-called floppy modes, i.e. zero-energy modes, besides rigid translations, as was already observed in Sec.~\ref{sec:energy_matching}.
However, compressive/tensile prestress may induce in the lattice periodic zero-energy modes, associated to buckling shapes of a single isolated elastic link.
The tensile and the first two compressive buckling loads and shapes for a single link, both hinged and clamped, are shown in Fig.~\ref{fig:buckling_modes}.

When one of these lattice periodic microbifurcations is attained, the coefficient matrix $\trans{\bZ_0}\bK(\bP)\,\bZ_0$ in Eq.~(\ref{eq:equilibrium_internal_strain}) admits a non-trivial zero-energy mode (floppy mode), so that the displacement vector $\Tilde{\bq}^*$ tends to infinity. As a result, the incremental strain energy (\ref{eq:strain_energy_lattice}) becomes unbounded.
Furthermore, continuing in a radial stress path beyond the singularity, the effective material may recover SE, condition (\ref{eq:SE_effective_medium}), or even PD, condition (\ref{eq:PD_effective_medium}).
However, the grid of rods is subject to local instabilities characterized by a periodic bifurcation mode, as illustrated in Fig.~\ref{fig:restabilization}.
In these conditions the homogenization framework does not capture the real behavior of the grid of rods, which remains unstable after the first global bifurcation corresponding to loss of SE in the effective material.

Fig.~\ref{fig:buckling_modes} reports the local modes of bifurcation occurring in all the rods, as they were isolated from each other, at different re-stabilization points (indicated  with the letters (\subref{fig:restabilization_mode_1})--(\subref{fig:restabilization_mode_7}) in panel (\subref{fig:domain_restabilization}) and referring to corresponding panels denoted with the same letters).
The square zones highlighted in orange in the figure represent the element of the homogenized continuum, which is left undeformed by the deformation mode corresponding to the local bifurcation mode shown to occur in the lattice.
This incremental deformation in the grid is therefore `invisible' to the effective continuum.
%

\section{Concluding remarks}
\label{sec:concluding}
Harnessing periodic lattices of elastic rods to design architected materials may lead to the erroneous conclusion that the latter are characterized by an ellipticity domain unbounded in tension.
As a consequence, it may be believed that these materials cannot fail under an ellipticity loss when these stress states prevail.
How these materials can be created to achieve a bounded stability domain has been shown in the present article through the use of sliders, namely, constraints allowing only relative sliding between two connected pieces of rod.
This result shows that homogenization leads to \textit{bounded stability domains}.
Moreover, our results open new possibilities for the realization of artificial materials with tunable properties and exhibiting strain localization within the elastic regime and for all possible directions in the stress space.

\section*{Acknowledgements}
G.B. and A.P. gratefully acknowledges the funding from the European Union’s Horizon 2020 research and innovation programme under the Marie Sklodowska-Curie grant agreement No 955944-REFRACTURE2. D.B. acknowledges financial support from ERC-ADG-2021-101052956-BEYOND.

\printbibliography

\appendix

\section{Full expression for the effective constitutive tensor}
\label{sec:effective_constitutive_tensor}
The complete analytic expression for the effective constitutive tensor of the lattice analyzed in Section~\ref{sec:grid} is here reported in dimensionless form $\fC =  \frac{A}{l} \, \Bar{\fC}$ (null components are omitted).

\begin{dgroup*}[style={\footnotesize},breakdepth={20}]
    \begin{dmath*}
        \Bar{\fC}_{1111} =\xi^{-1} \,,
    \end{dmath*}
    \begin{dmath*}
        \Bar{\fC}_{2222} =\chi \,,
    \end{dmath*}
    \begin{dmath*}
        \Bar{\fC}_{1212} =
        p_2^{3/2} \chi \left(-2 \lambda_1^2 \xi  \left(e^{\sqrt{p_1}}+1\right) p_1^{3/2} \sqrt{p_2} \chi  \left(\left(e^{\sqrt{p_2}}+1\right) p_2^{3/2}-2 \kappa_2 \left(e^{\sqrt{p_2}}-1\right)\right)+\\
        +\lambda_2^2 \left(e^{\sqrt{p_1}}+1\right) p_1^{5/2} \left(\left(e^{\sqrt{p_2}}-1\right) p_2^{3/2}-2 \kappa_2 \left(e^{\sqrt{p_2}}+1\right)\right)+\lambda_1^2 \xi\left(e^{\sqrt{p_1}}-1\right) p_1^2 \sqrt{p_2} \chi \left(\left(e^{\sqrt{p_2}}+1\right) p_2^{3/2}-2 \kappa_2 \left(e^{\sqrt{p_2}}-1\right)\right)+\\
        -2 \kappa_1 \lambda_1^2 \xi \left(e^{\sqrt{p_1}}+1\right) \sqrt{p_1} \sqrt{p_2} \chi  \left(\left(e^{\sqrt{p_2}}+1\right) p_2^{3/2}-2 \kappa_2 \left(e^{\sqrt{p_2}}-1\right)\right)+4 \kappa_1 \lambda_1^2 \xi  \left(e^{\sqrt{p_1}}-1\right) \sqrt{p_2} \chi  \left(\left(e^{\sqrt{p_2}}+1\right) p_2^{3/2}+\\
        -2 \kappa_2 \left(e^{\sqrt{p_2}}-1\right)\right)-2 \kappa_1 \lambda_2^2 \left(e^{\sqrt{p_1}}-1\right) p_1 \left(\left(e^{\sqrt{p_2}}-1\right) p_2^{3/2}-2 \kappa_2 \left(e^{\sqrt{p_2}}+1\right)\right)\right) / \left( \lambda_2^2\,D \right)
        \,,
    \end{dmath*}
    \begin{dmath*}
        \Bar{\fC}_{1221} = \Bar{\fC}_{2112} =
        2 p_1 p_2 \chi  \left(\left(e^{\sqrt{p_1}}+1\right) p_1^{3/2}-2 \kappa_1 \left(e^{\sqrt{p_1}}-1\right)\right) \left(\left(e^{\sqrt{p_2}}+1\right) p_2^{3/2}-2 \kappa_2 \left(e^{\sqrt{p_2}}-1\right)\right) / D
        \,,
    \end{dmath*}
    \begin{dmath*}
        \Bar{\fC}_{2121} =
        \left(-2 \kappa_1 \lambda_1^2 \xi  \left(e^{\sqrt{p_1}}+1\right) p_1^{3/2} p_2 \chi  \left(\left(e^{\sqrt{p_2}}+1\right) p_2^{3/2}-2 \kappa_2 \left(e^{\sqrt{p_2}}-1\right)\right)+\\
        +\lambda_2^2 \left(e^{\sqrt{p_1}}+1\right) p_1^{7/2} \left(-2 \left(e^{\sqrt{p_2}}+1\right) p_2^{3/2}+\left(e^{\sqrt{p_2}}-1\right) p_2^2+4 \kappa_2 \left(e^{\sqrt{p_2}}-1\right)+
        -2 \kappa_2 \left(e^{\sqrt{p_2}}+1\right) \sqrt{p_2}\right)+\\
        +\lambda_1^2 \xi  \left(e^{\sqrt{p_1}}-1\right) p_1^3 p_2 \chi  \left(\left(e^{\sqrt{p_2}}+1\right) p_2^{3/2}-2 \kappa_2 \left(e^{\sqrt{p_2}}-1\right)\right)
        -2 \kappa_1 \lambda_2^2 \left(e^{\sqrt{p_1}}-1\right) p_1^2 \left(-2 \left(e^{\sqrt{p_2}}+1\right) p_2^{3/2}+\\
        +\left(e^{\sqrt{p_2}}-1\right) p_2^2+4 \kappa_2 \left(e^{\sqrt{p_2}}-1\right)-2 \kappa_2 \left(e^{\sqrt{p_2}}+1\right) \sqrt{p_2}\right)\right) / \left( \lambda_1^2 \xi \,D \right)
        \,,
    \end{dmath*}
\end{dgroup*}
where the coefficient $D$ is given by
\begin{dmath*}[style={\footnotesize}]
    D=
    -2 \lambda_1^2 \xi  \left(e^{\sqrt{p_1}}+1\right) p_1^{3/2} p_2 \chi  \left(\left(e^{\sqrt{p_2}}+1\right) p_2^{3/2}-2 \kappa_2 \left(e^{\sqrt{p_2}}-1\right)\right)+\lambda_2^2 \left(e^{\sqrt{p_1}}+1\right) p_1^{5/2} \left(-2 \left(e^{\sqrt{p_2}}+1\right) p_2^{3/2}+\left(e^{\sqrt{p_2}}-1\right) p_2^2+4 \kappa_2 \left(e^{\sqrt{p_2}}-1\right)-2 \kappa_2 \left(e^{\sqrt{p_2}}+1\right) \sqrt{p_2}\right)+\\
    +\lambda_1^2 \xi  \left(e^{\sqrt{p_1}}-1\right) p_1^2 p_2 \chi  \left(\left(e^{\sqrt{p_2}}+1\right) p_2^{3/2}-2 \kappa_2 \left(e^{\sqrt{p_2}}-1\right)\right)+4 \kappa_1 \lambda_1^2 \xi  \left(e^{\sqrt{p_1}}-1\right) p_2 \chi  \left(\left(e^{\sqrt{p_2}}+1\right) p_2^{3/2}-2 \kappa_2 \left(e^{\sqrt{p_2}}-1\right)\right)+\\
    -2 \kappa_1 \lambda_1^2 \xi  \left(e^{\sqrt{p_1}}+1\right) \sqrt{p_1} p_2 \chi  \left(\left(e^{\sqrt{p_2}}+1\right) p_2^{3/2}-2 \kappa_2 \left(e^{\sqrt{p_2}}-1\right)\right)+\\
    -2 \kappa_1 \lambda_2^2 \left(e^{\sqrt{p_1}}-1\right) p_1 \left(-2 \left(e^{\sqrt{p_2}}+1\right) p_2^{3/2}
    +\left(e^{\sqrt{p_2}}-1\right) p_2^2+4 \kappa_2 \left(e^{\sqrt{p_2}}-1\right)-2 \kappa_2 \left(e^{\sqrt{p_2}}+1\right) \sqrt{p_2}\right) \,.
\end{dmath*}

\end{document}

%% file: definitions.tex
\DeclarePairedDelimiter{\jump}{\llbracket}{\rrbracket}

\newcommand{\scalp}{\bm \cdot}
\newcommand{\trans}[1]{#1^{\mathsf{T}}}

\newcommand{\deriv}[2]{\frac{\partial #1}{\partial #2}}

\newcommand{\bzero}{\bm 0}
\newcommand{\ba}{\bm a}
\newcommand{\bb}{\bm b}

\newcommand{\be}{\bm e}
\newcommand{\bef}{\bm f}
\newcommand{\bg}{\bm g}

\newcommand{\bn}{\bm n}

\newcommand{\bp}{\bm p}
\newcommand{\bq}{\bm q}

\newcommand{\bt}{\bm t}
\newcommand{\bu}{\bm u}

\newcommand{\bx}{\bm x}

\newcommand{\bA}{\bm A}
\newcommand{\bB}{\bm B}
\newcommand{\bC}{\bm C}
\newcommand{\bD}{\bm D}
\newcommand{\bE}{\bm E}

\newcommand{\bI}{\bm I}

\newcommand{\bK}{\bm K}
\newcommand{\bL}{\bm L}

\newcommand{\bN}{\bm N}

\newcommand{\bP}{\bm P}

\newcommand{\bS}{\bm S}
\newcommand{\bT}{\bm T}

\newcommand{\bW}{\bm W}

\newcommand{\bZ}{\bm Z}

\newcommand{\fC}{\mathbb C}

\newcommand{\fE}{\mathbb E}

\newcommand{\mB}{\mathcal B}
\newcommand{\mC}{\mathcal C}

\newcommand{\mE}{\mathcal E}

\newcommand{\mS}{\mathcal S}

\newcommand{\mV}{\mathcal V}
\newcommand{\mW}{\mathcal W}

%% file: sliding_grid.bib
@article{avazmohammadi_2016,
    title      = {Macroscopic Constitutive Relations for Elastomers Reinforced with Short Aligned Fibers: {{Instabilities}} and Post-Bifurcation Response},
    shorttitle = {Macroscopic Constitutive Relations for Elastomers Reinforced with Short Aligned Fibers},
    author     = {Avazmohammadi, Reza and Ponte Casta{\~n}eda, Pedro},
    year       = {2016},
    journal    = {J. Mech. Phys. Solids},
    series     = {{{SI}}:{{Pierre Suquet Symposium}}},
    volume     = {97},
    pages      = {37--67},
    issn       = {0022-5096},
    doi        = {10.1016/j.jmps.2015.07.007}
}

@book{bigoni_2012,
    title      = {Nonlinear Solid Mechanics: Bifurcation Theory and Material Instability},
    shorttitle = {Nonlinear Solid Mechanics},
    author     = {Bigoni, Davide},
    year       = {2012},
    publisher  = {{Cambridge University Press}},
    address    = {{Cambridge}},
    isbn       = {978-1-107-02541-7},
    langid     = {english},
    lccn       = {TA405 .B4983 2012}
}

@article{bigoni_2018,
    title   = {Bifurcation of Elastic Solids with Sliding Interfaces},
    author  = {Bigoni, D. and Bordignon, N. and Piccolroaz, A. and Stupkiewicz, S.},
    year    = {2018},
    journal = {Proc. R. Soc. Math. Phys. Eng. Sci.},
    volume  = {474},
    number  = {2209},
    pages   = {20170681},
    doi     = {10.1098/rspa.2017.0681}
}

@article{bordiga_2019,
    title   = {Prestress Tuning of Negative Refraction and Wave Channeling from Flexural Sources},
    author  = {Bordiga, G. and Cabras, L. and Piccolroaz, A. and Bigoni, D.},
    year    = {2019},
    journal = {Appl. Phys. Lett.},
    volume  = {114},
    number  = {4},
    pages   = {041901},
    issn    = {0003-6951},
    doi     = {10.1063/1.5084258},
    note    = {Promoted as Editor's Pick}
}

@article{bordiga_2019a,
    title      = {Free and Forced Wave Propagation in a {{Rayleigh-beam}} Grid: {{Flat}} Bands, {{Dirac}} Cones, and Vibration Localization vs Isotropization},
    shorttitle = {Free and Forced Wave Propagation in a {{Rayleigh-beam}} Grid},
    author     = {Bordiga, G. and Cabras, L. and Bigoni, D. and Piccolroaz, A.},
    year       = {2019},
    journal    = {Int. J. Solids Struct.},
    volume     = {161},
    pages      = {64--81},
    issn       = {0020-7683},
    doi        = {10.1016/j.ijsolstr.2018.11.007}
}

@article{bordiga_2021,
    title      = {Dynamics of Prestressed Elastic Lattices: {{Homogenization}}, Instabilities, and Strain Localization},
    shorttitle = {Dynamics of Prestressed Elastic Lattices},
    author     = {Bordiga, G. and Cabras, L. and Piccolroaz, A. and Bigoni, D.},
    year       = {2021},
    journal    = {J. Mech. Phys. Solids},
    volume     = {146},
    pages      = {104198},
    issn       = {0022-5096},
    doi        = {10.1016/j.jmps.2020.104198},
    langid     = {english}
}

@book{born_1955,
    title  = {Dynamical {{Theory}} of {{Crystal Lattices}}},
    author = {Born, Max and Huang, Kun},
    year   = {1955}
}

@article{carta_2017,
    title   = {``{{Deflecting}} Elastic Prism'' and Unidirectional Localisation for Waves in Chiral Elastic Systems},
    author  = {Carta, G. and Jones, I. S. and Movchan, N. V. and Movchan, A. B. and Nieves, M. J.},
    year    = {2017},
    journal = {Sci. Rep.},
    volume  = {7},
    number  = {1},
    pages   = {26},
    issn    = {2045-2322},
    doi     = {10.1038/s41598-017-00054-6},
    langid  = {english}
}

@article{cheng_2019,
    title      = {Micro/{{Nanoscale 3D Assembly}} by {{Rolling}}, {{Folding}}, {{Curving}}, and {{Buckling Approaches}}},
    author     = {Cheng, Xu and Zhang, Yihui},
    year       = {2019},
    journal    = {Adv. Mater.},
    volume     = {31},
    number     = {36},
    pages      = {1901895},
    issn       = {1521-4095},
    doi        = {10.1002/adma.201901895},
    langid     = {english}
}

@book{craster_2012,
    title       = {Acoustic {{Metamaterials}}: {{Negative Refraction}}, {{Imaging}}, {{Lensing}} and {{Cloaking}}},
    shorttitle  = {Acoustic {{Metamaterials}}},
    author      = {Craster, Richard V. and Guenneau, S{\'e}bastien},
    year        = {2012},
    publisher   = {{Springer Science \& Business Media}},
    googlebooks = {uv4lQ0tQJtkC},
    isbn        = {978-94-007-4813-2},
    langid      = {english}
}

@article{eremeyev_2020,
    title   = {Enriched Buckling for Beam-Lattice Metamaterials},
    author  = {Eremeyev, Victor A. and Turco, Emilio},
    year    = {2020},
    journal = {Mech. Res. Commun.},
    volume  = {103},
    pages   = {103458},
    issn    = {0093-6413},
    doi     = {10.1016/j.mechrescom.2019.103458},
    langid  = {english}
}

@article{furer_2018,
    title   = {Macroscopic Instabilities and Domain Formation in Neo-{{Hookean}} Laminates},
    author  = {Furer, J. and Ponte Casta{\~n}eda, P.},
    year    = {2018},
    journal = {Journal of the Mechanics and Physics of Solids},
    volume  = {118},
    pages   = {98--114},
    issn    = {0022-5096},
    doi     = {10.1016/j.jmps.2018.05.006},
    langid  = {english}
}

@article{garau_2019a,
    title      = {Transient Response of a Gyro-Elastic Structured Medium: {{Unidirectional}} Waveforms and Cloaking},
    shorttitle = {Transient Response of a Gyro-Elastic Structured Medium},
    author     = {Garau, M. and Nieves, M. J. and Carta, G. and Brun, M.},
    year       = {2019},
    journal    = {Int. J. Eng. Sci.},
    volume     = {143},
    pages      = {115--141},
    issn       = {0020-7225},
    doi        = {10.1016/j.ijengsci.2019.05.007},
    langid     = {english}
}

@article{guo_2018,
    title     = {Controlled Mechanical Assembly of Complex {{3D}} Mesostructures and Strain Sensors by Tensile Buckling},
    author    = {Guo, Xiaogang and Wang, Xueju and Ou, Dapeng and Ye, Jilong and Pang, Wenbo and Huang, Yonggang and Rogers, John A. and Zhang, Yihui},
    year      = {2018},
    journal   = {Npj Flex. Electron.},
    volume    = {2},
    number    = {1},
    pages     = {1--7},
    publisher = {{Nature Publishing Group}},
    issn      = {2397-4621},
    doi       = {10.1038/s41528-018-0028-y},
    langid    = {english}
}

@article{hill_1957,
    title   = {On Uniqueness and Stability in the Theory of Finite Elastic Strain},
    author  = {Hill, R.},
    year    = {1957},
    journal = {J. Mech. Phys. Solids},
    volume  = {5},
    number  = {4},
    pages   = {229--241},
    issn    = {0022-5096},
    doi     = {10.1016/0022-5096(57)90016-9}
}

@article{hill_1972,
    title   = {On Constitutive Macro-Variables for Heterogeneous Solids at Finite Strain},
    author  = {Hill, R.},
    year    = {1972},
    journal = {Proc. R. Soc. Lond. Math. Phys. Sci.},
    volume  = {326},
    number  = {1565},
    pages   = {131--147},
    doi     = {10.1098/rspa.1972.0001}
}

@article{hutchinson_2006,
    title   = {The Structural Performance of the Periodic Truss},
    author  = {Hutchinson, R.G. and Fleck, N.A.},
    year    = {2006},
    journal = {J. Mech. Phys. Solids},
    volume  = {54},
    number  = {4},
    pages   = {756--782},
    issn    = {00225096},
    doi     = {10.1016/j.jmps.2005.10.008},
    langid  = {english}
}

@article{madine_2021,
    title   = {Dynamic {{Green}}'s Functions in Discrete Flexural Systems},
    author  = {Madine, K H and Colquitt, D J},
    year    = {2021},
    journal = {Q. J. Mech. Appl. Math.},
    volume  = {74},
    number  = {3},
    pages   = {323--350},
    issn    = {0033-5614},
    doi     = {10.1093/qjmam/hbab006}
}

@article{mao_2018,
    title   = {Maxwell {{Lattices}} and {{Topological Mechanics}}},
    author  = {Mao, Xiaoming and Lubensky, Tom C.},
    year    = {2018},
    journal = {Annu. Rev. Condens. Matter Phys.},
    volume  = {9},
    number  = {1},
    pages   = {413--433},
    issn    = {1947-5454},
    doi     = {10.1146/annurev-conmatphys-033117-054235}
}

@article{mcphedran_2015,
    title   = {`{{Parabolic}}' Trapped Modes and Steered {{Dirac}} Cones in Platonic Crystals},
    author  = {McPhedran, R. C. and Movchan, A. B. and Movchan, N. V. and Brun, M. and Smith, M. J. A.},
    year    = {2015},
    journal = {Proc. Math. Phys. Eng. Sci. R. Soc.},
    volume  = {471},
    number  = {2177},
    issn    = {1364-5021},
    doi     = {10.1098/rspa.2014.0746},
    pmcid   = {PMC4984980},
    pmid    = {27547089}
}

@article{mishuris_2020,
    title     = {Waves in Elastic Bodies with Discrete and Continuous Dynamic Microstructure},
    author    = {Mishuris, Gennady S. and Movchan, Alexander B. and Slepyan, Leonid I.},
    year      = {2020},
    journal   = {Philos. Trans. R. Soc. Math. Phys. Eng. Sci.},
    volume    = {378},
    number    = {2162},
    pages     = {20190313},
    publisher = {{Royal Society}},
    doi       = {10.1098/rsta.2019.0313}
}

@article{mishuris_2020a,
    title      = {Localized Waves at a Line of Dynamic Inhomogeneities: {{General}} Considerations and Some Specific Problems},
    shorttitle = {Localized Waves at a Line of Dynamic Inhomogeneities},
    author     = {Mishuris, Gennady S. and Movchan, Alexander B. and Slepyan, Leonid I.},
    year       = {2020},
    journal    = {Journal of the Mechanics and Physics of Solids},
    volume     = {138},
    pages      = {103901},
    issn       = {0022-5096},
    doi        = {10.1016/j.jmps.2020.103901},
    langid     = {english}
}

@article{nestorovic_2004,
    title   = {Onset of Failure in Finitely Strained Layered Composites Subjected to Combined Normal and Shear Loading},
    author  = {Nestorovi{\'c}, M. D. and Triantafyllidis, N.},
    year    = {2004},
    journal = {J. Mech. Phys. Solids},
    volume  = {52},
    number  = {4},
    pages   = {941--974},
    issn    = {0022-5096},
    doi     = {10.1016/j.jmps.2003.06.001}
}

@article{nieves_2020,
    title      = {Rayleigh Waves in Micro-Structured Elastic Systems: {{Non-reciprocity}} and Energy Symmetry Breaking},
    shorttitle = {Rayleigh Waves in Micro-Structured Elastic Systems},
    author     = {Nieves, M. J. and Carta, G. and Pagneux, V. and Brun, M.},
    year       = {2020},
    journal    = {Int. J. Eng. Sci.},
    volume     = {156},
    pages      = {103365},
    issn       = {0020-7225},
    doi        = {10.1016/j.ijengsci.2020.103365},
    langid     = {english}
}

@article{nieves_2021,
    title   = {Directional {{Control}} of {{Rayleigh Wave Propagation}} in an {{Elastic Lattice}} by {{Gyroscopic Effects}}},
    author  = {Nieves, M. J. and Carta, G. and Pagneux, V. and Brun, M.},
    year    = {2021},
    journal = {Front. Mater.},
    volume  = {7},
    pages   = {422},
    issn    = {2296-8016},
    doi     = {10.3389/fmats.2020.602960}
}

@article{park_2019,
    title      = {Bio-{{Inspired Active Skins}} for {{Surface Morphing}}},
    author     = {Park, Yujin and Vella, Gianmarco and Loh, Kenneth J.},
    year       = {2019},
    journal    = {Sci. Rep.},
    volume     = {9},
    number     = {1},
    pages      = {18609},
    publisher  = {{Nature Publishing Group}},
    issn       = {2045-2322},
    doi        = {10.1038/s41598-019-55163-1},
    langid     = {english}
}

@article{pellegrino_1986,
    title   = {Matrix Analysis of Statically and Kinematically Indeterminate Frameworks},
    author  = {Pellegrino, S. and Calladine, C. R.},
    year    = {1986},
    journal = {Int. J. Solids Struct.},
    volume  = {22},
    number  = {4},
    pages   = {409--428},
    issn    = {0020-7683},
    doi     = {10.1016/0020-7683(86)90014-4}
}

@article{pellegrino_1990,
    title   = {Analysis of Prestressed Mechanisms},
    author  = {Pellegrino, S.},
    year    = {1990},
    journal = {Int. J. Solids Struct.},
    volume  = {26},
    number  = {12},
    pages   = {1329--1350},
    issn    = {0020-7683},
    doi     = {10.1016/0020-7683(90)90082-7}
}

@article{phani_2006,
    title   = {Wave Propagation in Two-Dimensional Periodic Lattices},
    author  = {Phani, A. Srikantha and Woodhouse, J. and Fleck, N. A.},
    year    = {2006},
    journal = {J. Acoust. Soc. Am.},
    volume  = {119},
    number  = {4},
    pages   = {1995--2005},
    issn    = {0001-4966},
    doi     = {10.1121/1.2179748},
    langid  = {english}
}

@article{piccolroaz_2020,
    title   = {Dynamic Phenomena and Crack Propagation in Dissimilar Elastic Lattices},
    author  = {Piccolroaz, A. and Gorbushin, N. and Mishuris, G. and Nieves, M.J.},
    year    = {2020},
    journal = {International Journal of Engineering Science},
    volume  = {149},
    pages   = {103208},
    issn    = {00207225},
    doi     = {10.1016/j.ijengsci.2019.103208},
    langid  = {english}
}

@article{pontecastaneda_1989,
    title   = {The Overall Constitutive Behaviour of Nonlinearly Elastic Composites},
    author  = {Ponte Casta{\~n}eda, P. and Spencer, Anthony James Merrill},
    year    = {1989},
    journal = {Proc. R. Soc. Lond. Math. Phys. Sci.},
    volume  = {422},
    number  = {1862},
    pages   = {147--171},
    doi     = {10.1098/rspa.1989.0023}
}

@article{pontecastaneda_1996,
    title   = {Exact Second-Order Estimates for the Effective Mechanical Properties of Nonlinear Composite Materials},
    author  = {Ponte Casta{\~n}eda, P.},
    year    = {1996},
    journal = {J. Mech. Phys. Solids},
    volume  = {44},
    number  = {6},
    pages   = {827--862},
    issn    = {0022-5096},
    doi     = {10.1016/0022-5096(96)00015-4}
}

@incollection{pontecastaneda_1997,
    title     = {Nonlinear {{Composites}}},
    booktitle = {Advances in {{Applied Mechanics}}},
    author    = {Ponte Casta{\~n}eda, P. and Suquet, Pierre},
    editor    = {{van der Giessen}, Erik and Wu, Theodore Y.},
    year      = {1997},
    volume    = {34},
    pages     = {171--302},
    publisher = {{Elsevier}},
    doi       = {10.1016/S0065-2156(08)70321-1}
}

@book{sanchez-palencia_1987,
    title     = {Homogenization {{Techniques}} for {{Composite Media}}},
    editor    = {{Sanchez-Palencia}, Enrique and Zaoui, Andr{\'e}},
    year      = {1987},
    series    = {Lecture {{Notes}} in {{Physics}}},
    volume    = {272},
    publisher = {{Springer Berlin Heidelberg}},
    address   = {{Berlin, Heidelberg}},
    doi       = {10.1007/3-540-17616-0},
    isbn      = {978-3-540-17616-9 978-3-540-47720-4},
    langid    = {english}
}

@article{santisidavila_2016,
    title   = {Localization of Deformation and Loss of Macroscopic Ellipticity in Microstructured Solids},
    author  = {{Santisi d'Avila}, M. P. and Triantafyllidis, N. and Wen, G.},
    year    = {2016},
    journal = {J. Mech. Phys. Solids},
    series  = {{{SI}}:{{Pierre Suquet Symposium}}},
    volume  = {97},
    pages   = {275--298},
    issn    = {0022-5096},
    doi     = {10.1016/j.jmps.2016.07.009}
}

@article{torrent_2013,
    title      = {Elastic Analog of Graphene: {{Dirac}} Cones and Edge States for Flexural Waves in Thin Plates},
    shorttitle = {Elastic Analog of Graphene},
    author     = {Torrent, Daniel and Mayou, Didier and {S{\'a}nchez-Dehesa}, Jos{\'e}},
    year       = {2013},
    journal    = {Phys. Rev. B},
    volume     = {87},
    number     = {11},
    pages      = {115143},
    doi        = {10.1103/PhysRevB.87.115143}
}

@article{triantafyllidis_1985,
    title   = {On the {{Comparison Between Microscopic}} and {{Macroscopic Instability Mechanisms}} in a {{Class}} of {{Fiber-Reinforced Composites}}},
    author  = {Triantafyllidis, N. and Maker, B. N.},
    year    = {1985},
    journal = {J. Appl. Mech.},
    volume  = {52},
    number  = {4},
    pages   = {794--800},
    issn    = {0021-8936},
    doi     = {10.1115/1.3169148}
}

@article{triantafyllidis_1993,
    title   = {Comparison of Microscopic and Macroscopic Instabilities in a Class of Two-Dimensional Periodic Composites},
    author  = {Triantafyllidis, Nicolas and Schnaidt, William C.},
    year    = {1993},
    journal = {J. Mech. Phys. Solids},
    volume  = {41},
    number  = {9},
    pages   = {1533--1565},
    issn    = {0022-5096},
    doi     = {10.1016/0022-5096(93)90039-I}
}

@book{willis_2002,
    title     = {Mechanics of Composites},
    author    = {Willis, John},
    year      = {2002},
    publisher = {{Ecole polytechnique, D\'epartement de m\'ecanique}}
}

@article{zaccaria_2011,
    title   = {Structures Buckling under Tensile Dead Load},
    author  = {Zaccaria, D. and Bigoni, D. and Noselli, G. and Misseroni, D.},
    year    = {2011},
    journal = {Proc. R. Soc. A},
    volume  = {467},
    number  = {2130},
    pages   = {1686--1700},
    doi     = {10.1098/rspa.2010.0505}
}

@article{zhang_2018,
    title   = {Fracturing of Topological {{Maxwell}} Lattices},
    author  = {Zhang, Leyou and Mao, Xiaoming},
    year    = {2018},
    journal = {New J. Phys.},
    volume  = {20},
    number  = {6},
    pages   = {063034},
    issn    = {1367-2630},
    doi     = {10.1088/1367-2630/aac765},
    langid  = {english}
}
